\documentclass[prl,aps,amsmath,amssymb]{revtex4-2}

\usepackage{graphicx,amssymb,amsmath}
\usepackage[utf8]{inputenc}
\usepackage{url,hyperref}	
\usepackage{bm}
\usepackage{color}
\usepackage{array}
\usepackage{enumerate}
\usepackage{booktabs}

\usepackage{theorem}
\usepackage[nodayofweek]{datetime}

\usepackage{hyperref}
\usepackage{etoolbox}

\usepackage{enumitem}
\usepackage{overpic}
\graphicspath{{./}{Figures/}}
\allowdisplaybreaks

\newenvironment{sciabstract}{%
\begin{quote} \bf}
{\end{quote}}

\usepackage{times}

\graphicspath{{./}{Figures/}}
\begin{document}

\title{An improved noise model for representing westerly wind bursts in the recharge oscillator model of ENSO}

\author{Georg A. Gottwald}
\affiliation{School of Mathematics and Statistics, University of Sydney, Australia}
%\email{georg.gottwald@sydney.edu.au} 

\author{Eli Tziperman}
\affiliation{Department of Earth and Planetary Sciences and School of Engineering and Applied Sciences, Harvard University, Cambridge, MA, USA}

\author{Alexey Fedorov}
\affiliation{Department of Earth and Planetary Sciences, Yale University, USA}

%\address{$^\star$School of Mathematics and Statistics, University of Sydney, Australia, georg.gottwald@sydney.edu.au}
%\address{$^\dagger$Department of Earth and Planetary Sciences and School of Engineering and Applied Sciences, Harvard University, Cambridge, MA, USA}
%\address{$^\ddagger$Department of Earth and Planetary Sciences, Yale University, USA}

\maketitle

%
%%%%%%%%%%%%%%%%%%%%%%%%%%%%%%%%%%%%%%%%%%%%%%%%
%
\begin{sciabstract}
Westerly wind bursts (WWBs) have long been known to have a major impact on the development of El Ni\~no events. In particular, they amplify these events, with stronger events associated with a higher number and stronger WWBs. We consider here a noise-driven recharge oscillator model of ENSO. Commonly, WWBs are represented by a state-dependent Gaussian noise that naturally reproduces the amplification of warm events. However, we show that many properties of WWBs and their effects on sea surface temperature (SST) are better captured by a conditional additive and multiplicative (CAM) noise, which presents a promising alternative to represent WWBs. In addition to recovering the sporadic nature of WWBs, CAM noise leads to an asymmetry between El Ni\~no and La Ni\~na events without the need for deterministic nonlinearities. Furthermore, CAM noise generates SST dynamics with a higher frequency of WWBs accompanying the largest events. This suggests that extreme warm events are better modelled by CAM noise. To cover the full spectrum of warm events, we propose a conditional noise model in which the wind stress is modelled by additive Gaussian noise for sufficiently small SSTs and by additive CAM noise once the SST exceeds a certain threshold. We show that this conditional noise model captures observed bulk statistical properties of ENSO equally well as the commonly used multiplicative Gaussian red noise model, but additionally better reproduces dynamical signatures such as the increased number of WWBs preceding large El Ni\~no events.
\end{sciabstract}
%
%%%%%%%%%%%%%%%%%%%%%%%%%%%%%%%%%%%%%%%%%%%%%%%%

%
%%%%%%%%%%%%%%%%%%%%%%%%%%%%%%%%%%%%%%%%%%%%%%%%
%

\section{Introduction} 

Westerly wind bursts (WWBs) in the equatorial Pacific last a week or two, have a longitudinal scale of a thousand km, and have preceded and amplified every major El Niño event in the observed record \citep{Harrison-Vecchi-1997:westerly, Lengaigne-Boulanger-Delecluse-et-al-2004:westerly, Yu-Fedorov-2022:essential}. WWBs trigger ocean Kelvin waves that accelerate the East Pacific warming \citep{Kessler-McPhaden-Weickmann-1995:forcing}. While these are weather events, they are not completely random, and tend to occur when the equatorial SST begins to warm \citep{Tziperman-Yu-2007:quantifying}, therefore amplifying developing El Niño events \citep{Eisenman-Yu-Tziperman-2005:westerly, Gebbie-Eisenman-Wittenberg-et-al-2007:modulation}. Further, WWBs occur more frequently during the active phase of the MJO \citep{Chiodi-Harrison-Vecchi-2014:subseasonal,Liang-Fedorov-2021:linking}. Because of this implied state-dependency, WWB events are commonly represented as a multiplicative Gaussian noise term in simplified ENSO models studying their role in El Niño events \citep[e.g.,][]{Perez-Moore-Zavala-Garay-et-al-2005:comparison, Vialard-Jin-McPhaden-et-al-2025:el}. In this formulation, the red noise amplitude is multiplied by a Heaviside function of the temperature and by the East Pacific temperature itself, restricting the events to warmer than normal temperatures, and making sure the stochastic noise amplitude increases with the SST anomaly, as motivated by observational analysis \citep{Tziperman-Yu-2007:quantifying}. This allows an excellent fit to the observed ENSO SST characteristics, including the spectrum and PDF \citep{Vialard-Jin-McPhaden-et-al-2025:el,Han-Fedorov-Vialard-2026:realistic}. 

\begin{figure}[!htp]
    \centering
    \includegraphics[width=0.45\linewidth]{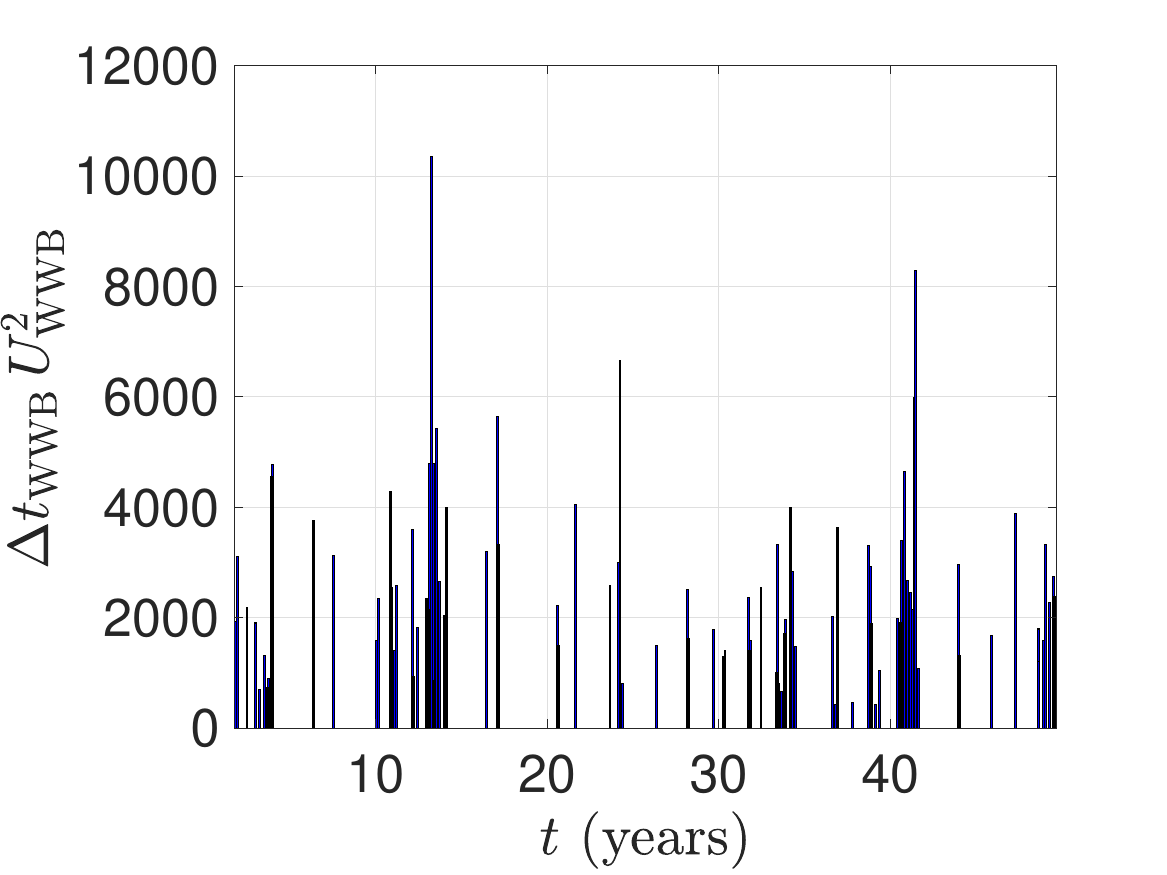}
    \includegraphics[width=0.45\linewidth]{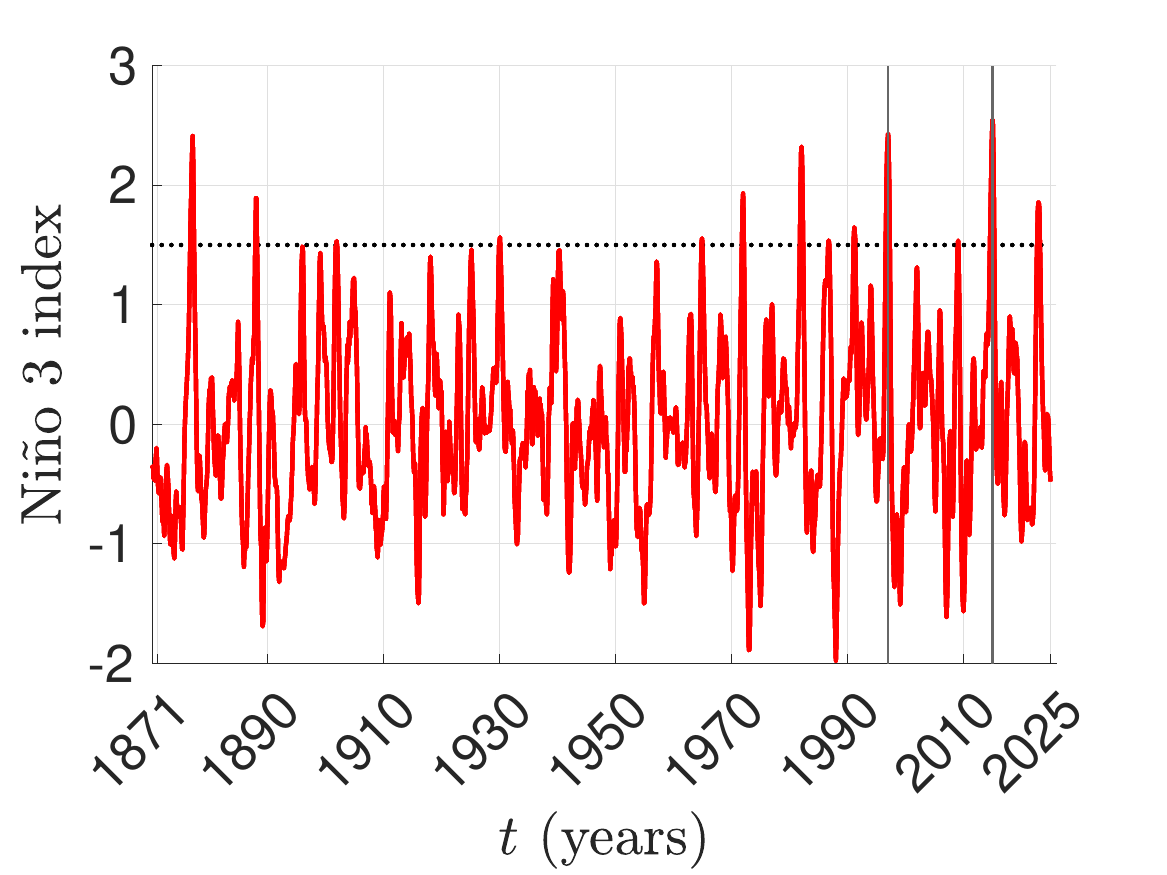}
    \caption{\textbf{WWBs from a climate model and NINO3 from observations.} (a) An inferred measure of the time-integrated wind stress due to westerly wind bursts in a global climate model \cite[CESM2,][]{Danabasoglu-Lamarque-Bacmeister-et-al-2020:community}. Shown is $\Delta t_{\rm{WWB}}\,  U_{\rm{WWB}}^2$, where the WWB duration is given by $\Delta t_{\rm{WWB}}$ (in days) and the surface wind speed strength is $U_{\rm{WWB}}$ (in $m/s$).
    (b) Observed NINO3 index from 1870 until 2026 \citep{Rayner-Parker-Horton-et-al-2003:global}. A moving average over 4 months was applied to observations. The horizontal line demarcates an index of 1.5, which we use to separate large El Niño events from normal ones. The vertical lines denote the years 1997 and 2015, for which the large El Niño events were accompanied by strong WWBs.}
\label{fig:obs}
\end{figure}

In this work, we examine an alternative to this stochastic forcing formulation, which provides an improved representation of the effects of WWBs on ENSO. We show that while Gaussian noise reproduces the global statistical features of ENSO, an alternative formulation is required to better represent the character of WWBs. To motivate our proposed noise model, consider Figure~\ref{fig:obs}a, which shows an inferred measure of the time-integrated wind stress associated with WWBs from a numerical simulation of a global configuration of the Community Atmospheric Model~\citep{Conley-Garcia-Kinnison-et-al-2012:description}. \cite{Lian-Tang-Zhou-et-al-2018:westerly} show that even an earlier version of the Community Atmospheric Model simulates WWBs quite realistically. Specifically, we show $\Delta t_{\rm{WWB}}\,  U_{\rm{WWB}}^2$, where $\Delta t_{\rm{WWB}}$ is the WWB duration and $U_{\rm{WWB}}$ is the surface wind speed strength. This measure attempts to represent the effect of these events on the ocean, which depends on both their amplitude and their duration. The observed time series in Figure~\ref{fig:obs}a with its sporadic high-amplitude peaks resembles that of so-called correlated additive and multiplicative (CAM) noise, which has been used to study non-Gaussian oceanic SST anomalies and atmospheric dynamics \citep{Sura-Sardeshmukh-2008:global, Sardeshmukh-Sura-2009:reconciling, Penland-Sardeshmukh-2012:alternative, Sardeshmukh-Penland-2015:understanding}. Numerical simulations suggest that resolving the sporadic peaky nature of WWBs is crucial in modelling ENSO \citep{Yu-Fedorov-2020:role}. This suggests replacing the commonly employed Gaussian stochastic forcing used to represent WWBs in idealized ENSO models with non-Gaussian CAM noise. As we will show, such CAM noise also naturally reproduces the observed asymmetry between El Ni\~no and La Ni\~na events, without the inclusion of any deterministic nonlinearities designed to promote such asymmetry. Using CAM noise to explain the observed asymmetry in amplitude and persistence of El Niño and La Niña events has also been suggested by \cite{Martinez-Villalobos-Newman-Vimont-et-al-2019:observed}, who showed that the asymmetry can be captured by a multivariate linear model driven by CAM noise. Moreover, we will show that CAM noise can generate characteristic dynamical signatures of major El Ni\~no events. In particular, major El Ni\~no events are the result of a sustained driving by several large amplitude WWBs \citep{Chiodi-Harrison-Vecchi-2014:subseasonal,Liang-Fedorov-2021:linking}. We will show that a recharge oscillator (RO) model driven by CAM noise well reproduces this dynamical feature, whereas standard multiplicative Gaussian noise fails to capture it. 

We wish to make the case that CAM noise is a more appropriate noise model to model WWBs that are known to occur with a developing El Niño warming, further amplifying large El Ni\~no events \citep{Yu-Weller-Liu-2003:case, Tziperman-Yu-2007:quantifying, LiangFedorov2021CD}. Stochastic wind forcing over cooler oceanic environments may be more appropriately captured by Gaussian noise. We therefore propose a conditional noise model, that we will coin CON, in which the noise is Gaussian unless the SST exceeds a certain threshold when the noise switches to CAM noise. This conditional noise model will be shown to well reproduce the observed signatures of large El Niño events, while preserving the overall observed statistical properties of ENSO.

The paper is organized as follows. In Section~\ref{sec:model} we introduce the recharge oscillator model \citep{Jin-1997:equatorial-I, Burgers-Jin-Oldenborgh-2005:simplest, Vialard-Jin-McPhaden-et-al-2025:el}, including a deterministic nonlinearity representing an asymmetric response of SST to changes in the thermocline. Section~\ref{sec:noise} introduces the stochastic forcing. We discuss both additive and multiplicative noise and introduce a coloured Gaussian noise model as well as CAM noise. We demonstrate that coloured Gaussian noise gives rise to Brownian motion when integrated, whereas the CAM noise gives rise to L\'evy noise with abrupt jumps, which are more similar to WWBs and have the potential to trigger large El Ni\~no events. Section~\ref{sec:results} numerically explores the effect of three noise (stochastic forcing) scenarios for representing WWBs in an RO model: (1) a standard Gaussian noise used as multiplicative noise based on the East Pacific SST, representing what is the emerging consensus in the ENSO literature~\citep{Perez-Moore-Zavala-Garay-et-al-2005:comparison, Vialard-Jin-McPhaden-et-al-2025:el}; (2) a CAM noise, and finally, (3) a conditional noise that alternates between Gaussian noise for cold SST anomalies and CAM noise for positive anomalies. We argue that the conditional CAM noise may better capture the dynamics of large El Ni\~no events. We conclude in Section~\ref{sec:disc} with a discussion of our results. 

%
%%%%%%%%%%%%%%%%%%%%%%%%%%%%%%%%%%%%%%%%%%%%%%%%
%

\section{Methods}
\label{sec:methods}

\subsection{The recharge oscillator model} 
\label{sec:model}

We consider the recharge oscillator (RO) model \citep{Jin-1997:equatorial-I,Burgers-Jin-Oldenborgh-2005:simplest,Jin-Lin-Timmermann-et-al-2007:ensemble,Vialard-Jin-McPhaden-et-al-2025:el} for the Western Pacific thermocline depth anomaly $h_w$ and the Eastern Pacific sea-surface temperature (SST) anomaly $T_e$, 
\begin{align}
    {\dot T}_e &= -r(T_e-\gamma h_e(t))
    \label{eq:Te0} \\
    {\dot h}_w &= -\varepsilon(h_w+ a\tau(t)),
    \label{eq:hw0} 
\end{align}
where the Eastern Pacific thermocline depth is expressed as 
\begin{align}
 h_e = h_w+ d \, \tau,
 \label{eq:he}
\end{align}
with non-dimensional wind stress 
\begin{align}
    \tau = b T_e + \nu \xi(t).
    \label{eq:tau}
\end{align}
We set $d=1\, m^3/N$ throughout. The parameters $r$ and $\varepsilon$ control the characteristic decay time scales for the SST and for the thermocline depth, respectively, and $a$, $b$ are positive constants. The noise $\xi(t)$ with amplitude $\nu>0$ represents the effect of westerly wind bursts \citep{Vialard-Jin-McPhaden-et-al-2025:el}. For westerly wind anomalies, $\xi>0$, the noise acts to reduce the West Pacific thermocline depth $h_w$ and amplifies the East Pacific SST $T_e$, and vice versa for $\xi<0$.
The RO model can be concisely written as 
\begin{align}
     \dot{T}_e&= R T_e + F_1 h_w + \sigma_T \, \xi(t)
    \label{eq:Te} \\
     \dot{h}_w&= - F_2 T_e - \varepsilon h_w  - \sigma_h \, \xi(t),
    \label{eq:hw} 
    %% \nu <-- F_1 \nu !
\end{align}
where $R=r(\gamma bd-1)$ represents the Bjerknes feedback and $F_1=r\gamma$ and $F_2=ab\varepsilon$ are constants, and $\sigma_T = F_1 d\nu$ and $\sigma_h = F_1F_2 d \nu/(R+r)$ denote the amplitudes of the noise \citep{Vialard-Jin-McPhaden-et-al-2025:el}. Without stochastic driving (i.e., $\nu=0$) the dynamics of (\ref{eq:Te0})--(\ref{eq:hw0}) is that of a linear oscillator with frequency $\omega^2 = F_1F_2-(\varepsilon+R)^2/4$ and a growth/damping rate $\lambda = (\varepsilon-R)/2$.

We choose parameters such that one time unit in (\ref{eq:Te})--(\ref{eq:hw}) corresponds to 1 month. We remark that an oversimplified fit of the RO model to match observations or climate model simulation of ENSO leads to a biased set of parameters that shows ENSO to be damped~\citep{Weeks-Tziperman-2025:challenges} even if it is self-sustained (positive growth rate), although that should not affect our focus here. 

To account for the observed asymmetry that El Niño events are typically of a larger magnitude than La Niña events, the RO model can be extended to include nonlinear terms in the temperature equation \cite[see][for a recent review]{Vialard-Jin-McPhaden-et-al-2025:el}. Here, we use in one of our three numerical experiments below a nonlinear response of the SST to the thermocline depth, motivated by the parameterization of subsurface temperature in the CZ model \citep{Zebiak-Cane-1987:model}. This involves employing a nonconstant response $\gamma$ of the SST to changes in the thermocline with
\begin{align}
    \gamma = 
    \begin{cases}
    \gamma_+ \quad\text{if } h_e\ge 0\\
    \gamma_- \quad\text{if } h_e< 0,
    \end{cases}.
    \label{eq:gamma}
\end{align} 
With $\gamma_+>\gamma_-$, equation \eqref{eq:Te0} implies a stronger effect for a positive thermocline depth anomaly on the SST than for a negative thermocline depth anomaly. This enhances El Niño amplitudes relative to those of La Niña events, consistent with observations. Alternatively, one could add quadratic nonlinearities such as $\beta T^2$, with $\beta > 0$, in the temperature equation \eqref{eq:Te0} to represent physical processes that favor the growth of El Niño relative to La Niña \cite[see, e.g.,][for a systematic investigation of the impact of different nonlinear terms]{Vialard-Jin-McPhaden-et-al-2025:el,Liu-Vialard-Fedorov-et-al-2024:why}. Besides such deterministic nonlinearities, state-dependent stochastic drivers \citep{Jin-Lin-Timmermann-et-al-2007:ensemble, Levine-Jin-McPhaden-2016:extreme, Levine-Jin-2017:simple} and non-Gaussian noise \citep{Bianucci-2016:analytical, Bianucci-Capotondi-Merlino-et-al-2018:estimate} were proposed as further dynamic ingredients to account for the observed asymmetry. In the following, we introduce several prototypical noise models. In Section~\ref{sec:results} we will discuss their effect on the dynamics of ENSO, including its asymmetry, within the RO model, and then combine them to generate a conditional noise model that exhibits more realistic signatures consistent with observations. 

%
%%%%%%%%%%%%%%%%%%%%%%%%%%%%%%%%%%%%%%%%%%%%%%%%
%

\subsection{Noise models}
\label{sec:noise}

We consider both additive and multiplicative noise with
\begin{align}
    \nu = \nu_0+\nu_1{\rm{max}}(0,T_e). 
    \label{eq:noise_struc}
\end{align}
This form of the multiplicative noise, with $\nu_1\neq 0$, takes into account that atmospheric noise, such as WWBs, occurs more frequently and with higher amplitude over a warmer equatorial ocean. Such multiplicative noise naturally introduces an asymmetry with larger El Ni\~no events compared to La Ni\~na events \citep{Jin-Lin-Timmermann-et-al-2007:ensemble, Levine-Jin-McPhaden-2016:extreme,Han-Fedorov-Vialard-2026:realistic}.

Additive noise with $\nu_1=0$ cannot generate the desired ENSO asymmetry for a Gaussian driving noise such as coloured Ornstein-Uhlenbeck (OU) processes. However, as we will show, certain non-Gaussian noise models, such as correlated additive and multiplicative (CAM) noise, can generate the asymmetry even when only applied additively. CAM noise is defined as the stochastic process governed by \citep[e.g.,][]{Thompson-Kuske-Monahan-2017:reduced},
\begin{align}
    % d\xi = L \xi \, dt - \frac{E}{2} \xi\, dt + (E \xi+G) \circ dW_1 + B \,dW_2
    d\xi = c_1 \xi \, dt + (c_2 \xi+c_3) \circ dW_1 + c_4 \,dW_2.
    \label{eq:CAM}
\end{align}
% c1 = L
% c2 = E
% c3 = G
% c4 = B
Here, $W_{1,2}$ are independent Brownian motions (Wiener noise), with independent normally distributed random increments $dW_{1,2}$,  
%$W_i^{n+1}-W_i^{n}$, 
and the $\circ$-symbol denotes Stratonovich noise. This noise model naturally appears when modelling the effect of fast dynamic processes onto slower ones and has found numerous applications in atmospheric and climate dynamics~\citep{Sura-Sardeshmukh-2008:global, Sardeshmukh-Sura-2009:reconciling, Majda-Franzke-Crommelin-2009:normal, Penland-Sardeshmukh-2012:alternative, Sardeshmukh-Penland-2015:understanding, Gottwald-Crommelin-Franzke-2017:stochastic, Gottwald-2021:model} and in particular, in the context of ENSO \citep{Martinez-Villalobos-Newman-Vimont-et-al-2019:observed}.

The noise model \eqref{eq:CAM} contains both a purely Gaussian noise and a non-Gaussian noise as particular limits, depending on its parameters. For $c_2=c_3=0$ and $c_1<0$, we have
\begin{align}
    d\xi = -|c_1|\xi \, dt + c_4 \,dW_2,
    \label{eq:Ornstein-Uhlenbeck}
\end{align}
and the CAM noise reduces to an Ornstein-Uhlenbeck noise with an asymptotic Gaussian distribution and  zero mean~\citep{Pavliotis-Stuart-2008:multiscale}. The central limit theorem ensures that if a Gaussian process is integrated in time, the resulting random variables are also distributed according to a Gaussian distribution.

However, for $c_1<0$ and $c_2\neq 0$ CAM noise is non-Gaussian and lies in the domain of attraction of $\alpha$-stable processes \citep{Kuske-Keller-2001:rate}. This means that if such a process is integrated in time, it generates random variables that are drawn from an $\alpha$-stable distribution. Such $\alpha$-stable processes $L_{\alpha,\beta,\eta}$ are characterized by discrete jumps similarly to L\'evy processes, and are parametrized by three parameters $\alpha$, $\beta$ and $\eta$. The stability parameter $\alpha\in(0,2]$ determines the occurrence and size of the jumps. For $\alpha=2$ we obtain a continuous Gaussian process without any discrete jumps. For $\alpha<2$, however, the variance of such a process is not defined, as discrete jumps of arbitrary size have non-vanishing probability, and for $\alpha<1$ even the mean ceases to exist. The skewness parameter $\beta\in [-1,1]$ controls the direction of the jumps with $\beta=1$ allowing for only positive jumps, $\beta=-1$ allowing for only negative jumps, $\beta=0$ allowing, on average, for as many positive as negative jumps, and values of $\beta$ in between quantifying the probability of having positive or negative jumps. The scale parameter $\eta$ reduces to the variance for the Gaussian case with $\alpha=2$. For more details on $\alpha$-stable processes, we refer the reader to \cite{Applebaum-2009:levy, Chechkin-Metzler-Klafter-et-al-2008:introduction}. 

For $c_4\neq 0$ the mean of $\xi$ is well-defined, and one has explicit expressions for the parameters of the resulting L\'evy process $\alpha$, $\beta$ and $\eta$ as functions of the parameters of the CAM process \citep{Kuske-Keller-2001:rate,Thompson-Kuske-Monahan-2017:reduced}. For integrated CAM noise, the stability parameter $\alpha$ of the resulting $\alpha$-stable process $L_{\alpha,\eta,\beta}$ is given by 
\begin{align}
    \alpha&=-2{c_1}/{c_2^2},
    \label{eq:alphaCAM}
\end{align}
the skewness parameter is given by,
\begin{align}
    \beta=\tanh\left(\frac{\pi c_3(\alpha-1)}{2c_4}\right),
    \label{eq:betaCAM}
\end{align}
and the scale parameter $\eta$ is given by 
\begin{align*}
    \eta=\left(\frac{2\cosh\left(\frac{\pi c_3(\alpha-1)}{2c_4}\right)}{c_2^{\alpha+1}\alpha N} \Gamma(1-\alpha)\cos\left(\frac{\pi}{2}\alpha\right)\right)^\frac{1}{\alpha},
\end{align*}
with
\begin{align*} 
    N=2\pi(2c_4)^{-\alpha} \; 
    \frac{\Gamma(\alpha)}{c_2\Gamma(z)\Gamma(\bar z)}, 
\end{align*}
where the bar denotes the complex conjugate and 
\begin{align*} 
    z=\frac{\alpha+1}{2}+ i\frac{c_3(\alpha-1)}{c_4}.
\end{align*}

\begin{figure}
	\begin{overpic}[width=0.47\linewidth]{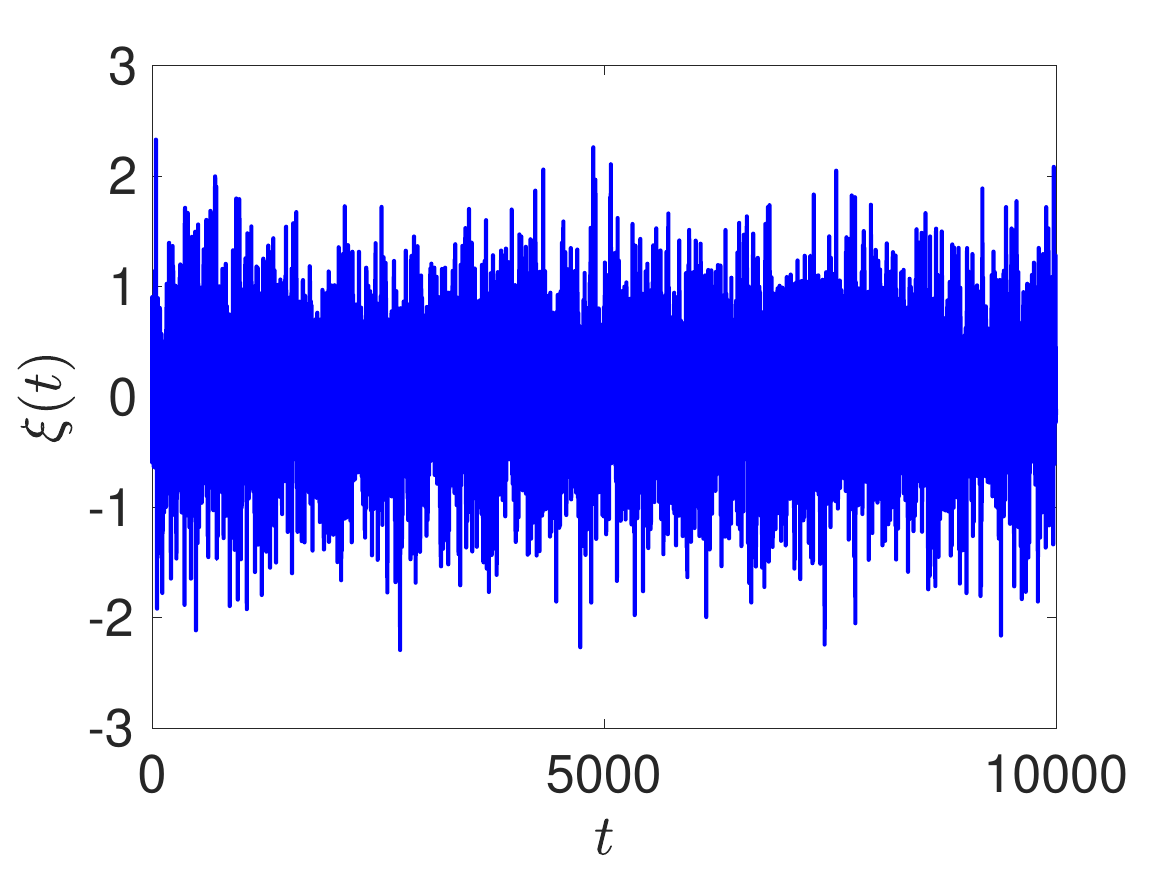}
    \put(14,73){\small (a)}
    \end{overpic}
	\begin{overpic}[width=0.47\linewidth]{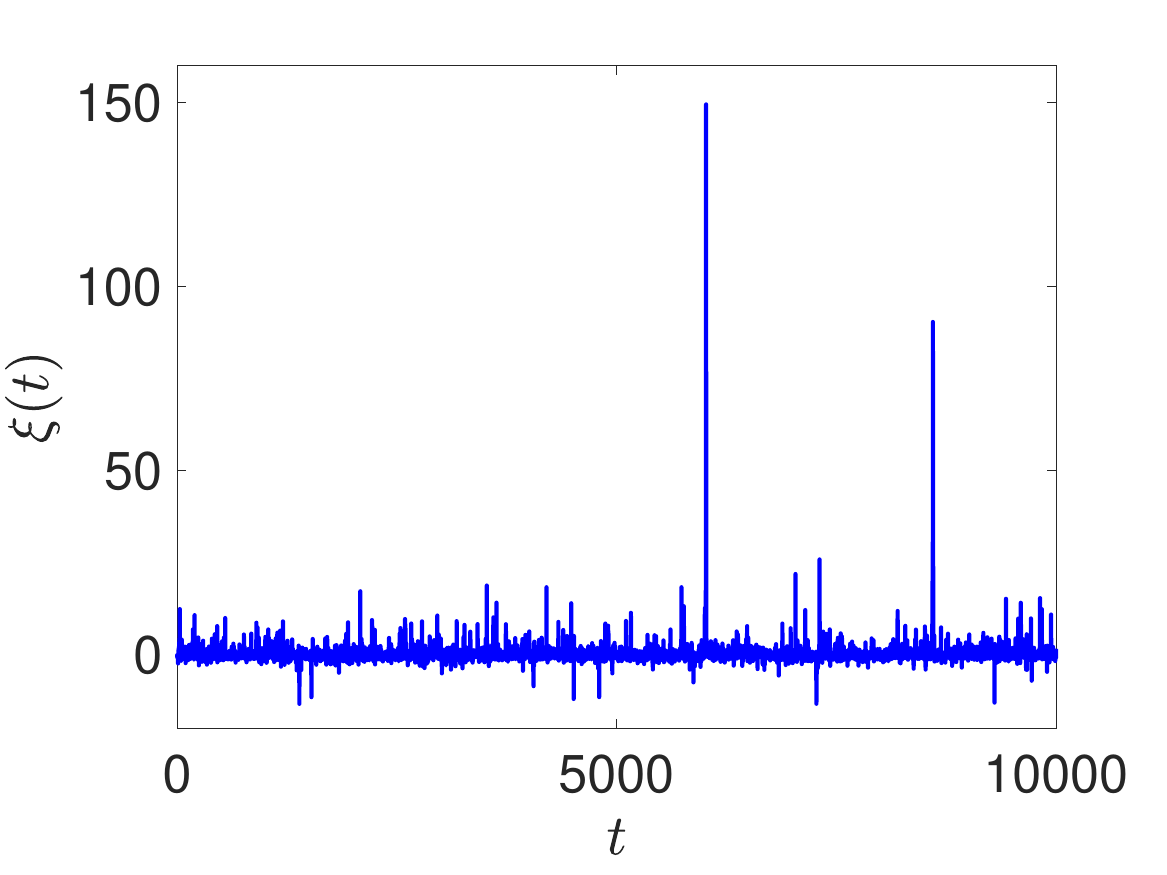}
    \put(14,73){\small (b)}
    \end{overpic}
	\begin{overpic}[width=0.47\linewidth]{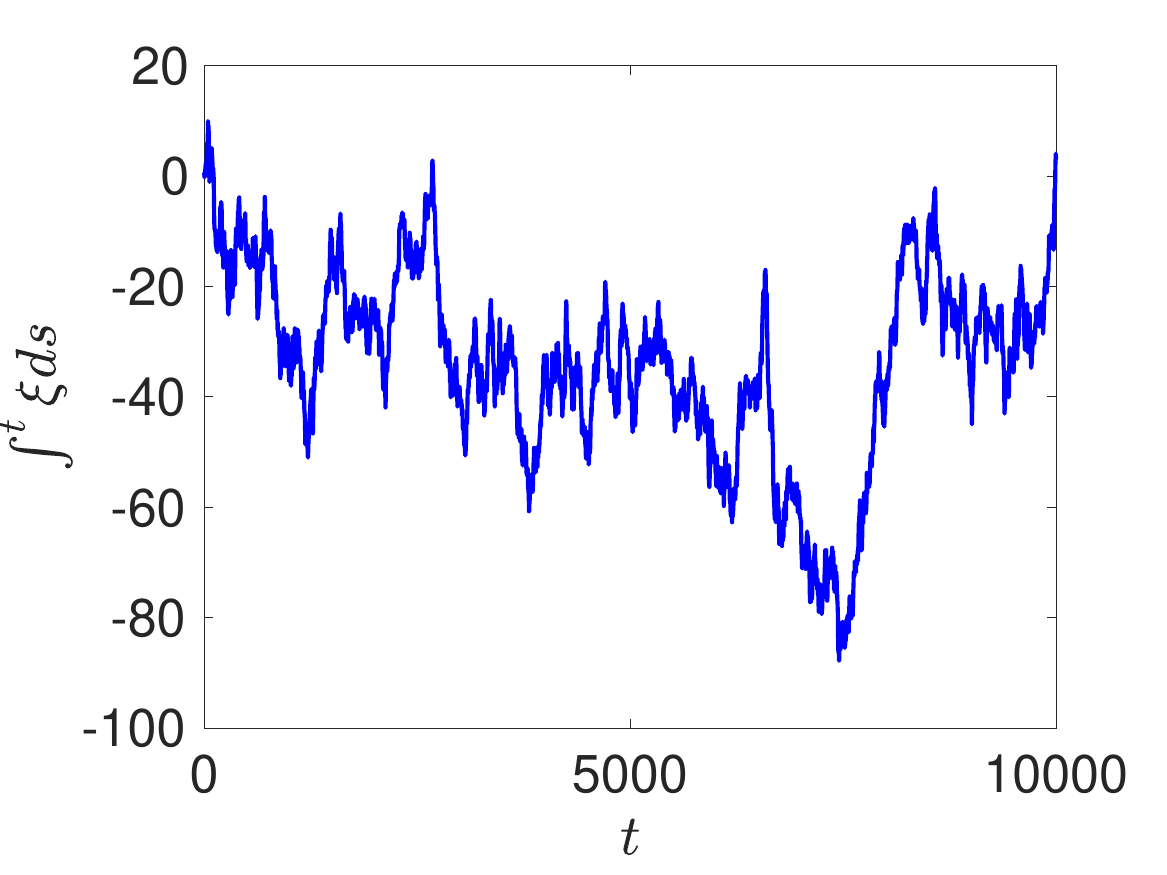}			
    \put(14,73){\small (c)}
    \end{overpic}
	\begin{overpic}[width=0.47\linewidth]{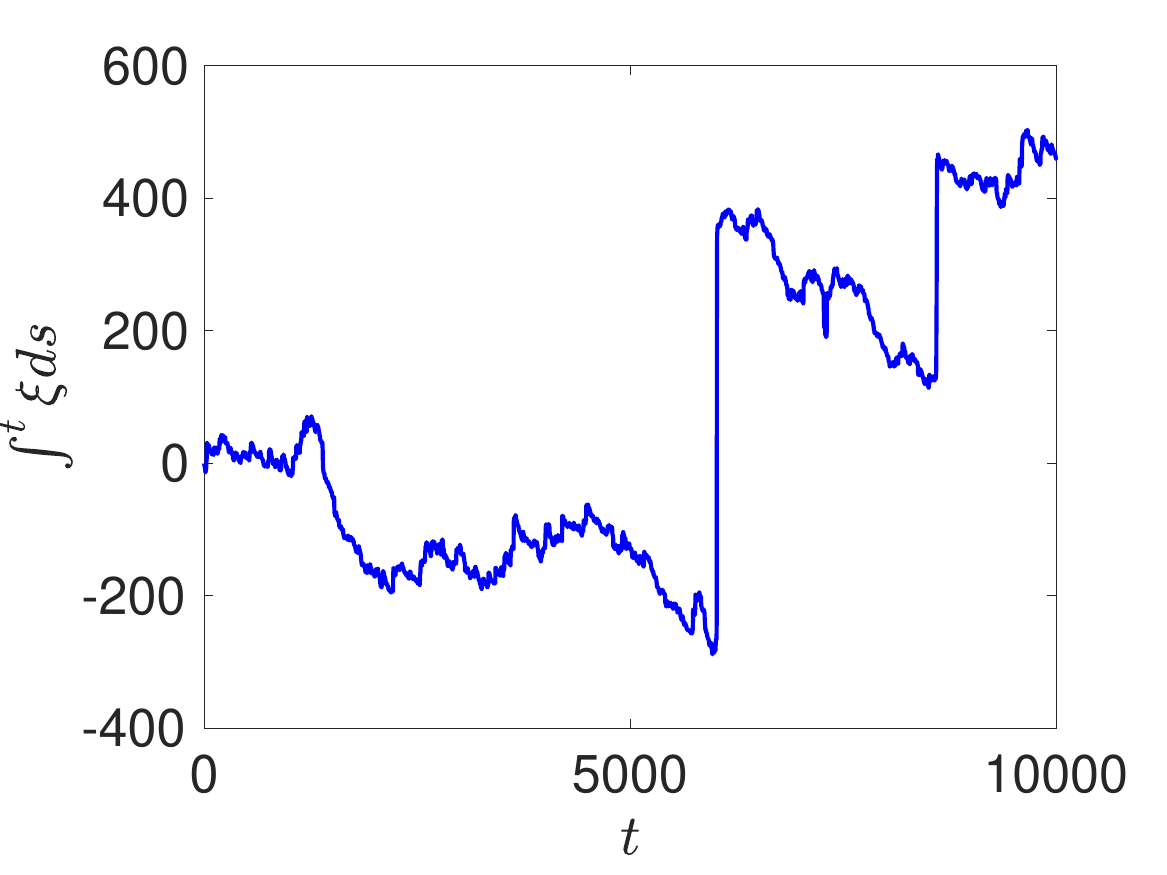}
    \put(14,73){\small (d)}
    \end{overpic}
	\caption{\textbf{Time series and integrated time series of OU and CAM noise processes.} (a) OU process with a characteristic decorrelation decay time of $4.9$ days ($c_1=-1.4$, $c_4=0.7$, $c_2=c_3=0$). (b)  CAM noise process with intermittent peaks ($c_1=1.22$, $c_2=1.14$, $c_3=0.65$ and $c_4=0.8$ with corresponding $\alpha=1.88$ and $\beta=0.81$). (c) and (d) show the integrals of (a) and (b), correspondingly.}
\label{fig:CAM}
\end{figure}

Figure~\ref{fig:CAM} shows examples of CAM processes $\xi(t)$ and their integrals $\Xi = \int^t \xi(s)ds$. Panel (a) depicts a Gaussian OU process with $c_1=-2/3$, $c_4=0.7$, $c_2=c_3=0$, which when integrated yields Brownian motion as shown in panel (c), in accordance with the central limit theorem. Panel (b) shows non-Gaussian CAM noise with intermittent unbounded peaks with $c_1=1.22$, $c_2=1.14$, $c_3=0.65$ and $c_4=0.8$, which, when integrated, leads to $\alpha$-stable noise with jumps, as shown in panel (d). In the following, we will call noise obtained from \eqref{eq:CAM} OU noise if $c_2=c_3=0$ and $c_1<0$. We will use the term CAM noise only for those processes \eqref{eq:CAM} that are not OU processes. For parameters of the CAM process that lead to $\alpha$-stable noise with only positive jumps, i.e., $\beta=1$, we see that jumps are caused by sporadic peaks of varying sizes of the CAM noise. These sporadic large-amplitude peaks will constitute our prototypical noise representation of WWBs (cf. Figure~\ref{fig:obs}a). In our application, we need to limit the magnitude of the CAM noise for WWBs not to have an arbitrarily large amplitude. We will therefore replace below the output of the CAM noise \eqref{eq:CAM} by ${\rm{max}}(\xi,\theta)$, with $\theta=3$ unless stated otherwise.

%
%
%%%%%%%%%%%%%%%%%%%%%%%%%%%%%%%%%%%%%%%%%%%%%%%%
%
\section{Results}
\label{sec:results}

To examine the effect on ENSO of the noise model as a representation of WWBs, we consider three different noise models: 
\begin{enumerate}[label=(\arabic*)]
    \item A multiplicative OU noise, a commonly used representation of WWBs \citep{Jin-Lin-Timmermann-et-al-2007:ensemble,Levine-Jin-McPhaden-2016:extreme,Vialard-Jin-McPhaden-et-al-2025:el}, denoted \textbf{OU}.
    
    \item An additive noise model with non-Gaussian CAM noise, denoted \textbf{CAM}.
    
    \item A conditional noise model, denoted \textbf{CON}, where we employ additive OU noise for negative SST anomaly, $T_e<0$, and then allow for more intense WWBS modeled by adding CAM noise for $T_e>0$.
\end{enumerate}

Note that in \textbf{CON}, the noise is state-dependent, and can therefore be considered multiplicative noise, even if its amplitude is not proportional to the temperature as in \textbf{OU}.

Figure~\ref{fig:obs}b shows the NINO3 index from 1870 until 2026 from the NOAA data set \citep{Rayner-Parker-Horton-et-al-2003:global}, where we subtracted the seasonal cycle and employed a moving average over a 4-month window. The focus of this work is large El Niño events, which we define to be those events with an NINO3 index larger than 1.5 °C. We expect such large events to be preceded by strong WWBs, consistent with observed large El Niño events such as those of 1997 and 2015 (marked by vertical gray lines in Figure~\ref{fig:obs}b), which were accompanied by multiple strong WWBs \citep{Puy-Vialard-Lengaigne-et-al-2019:influence}.

\begin{figure}
	\begin{overpic}[width=0.47\linewidth]{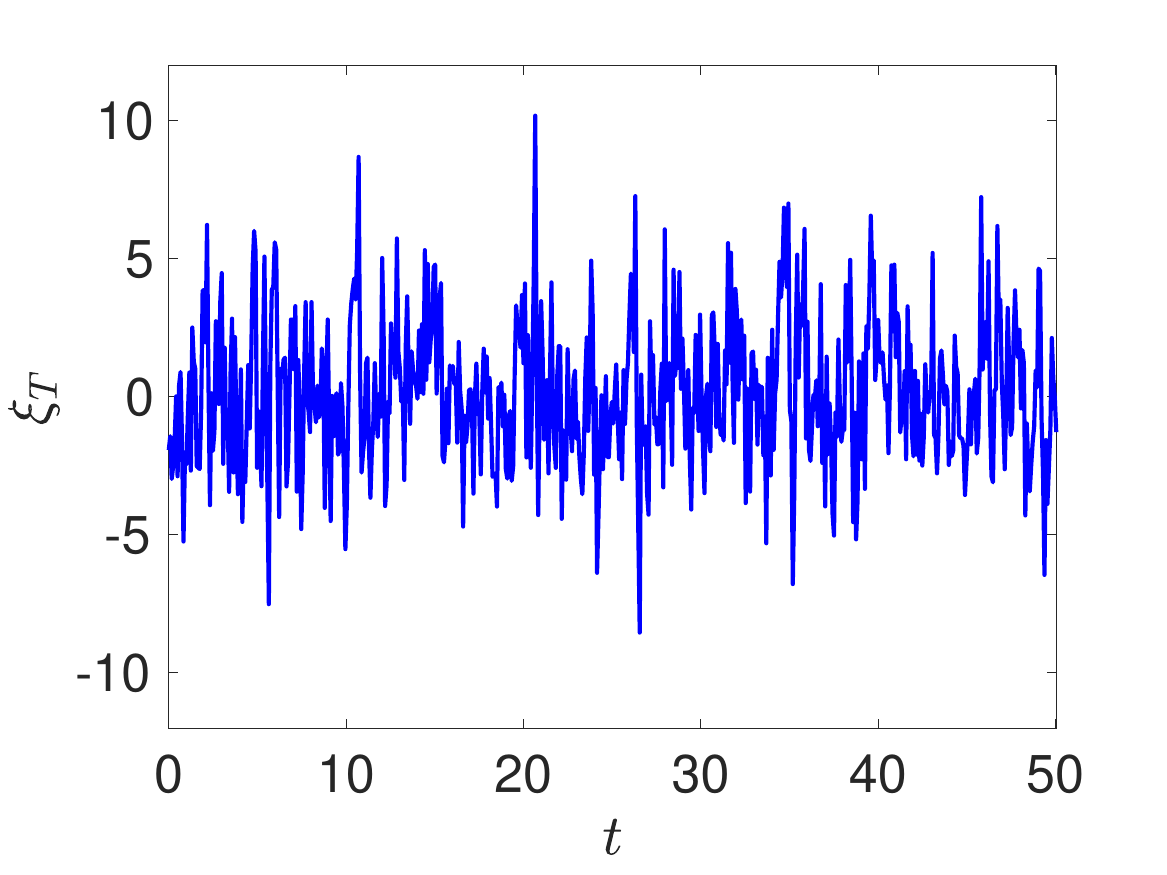}
    \put(14,73){\small (a)}
    \end{overpic}
	\begin{overpic}[width=0.47\linewidth]{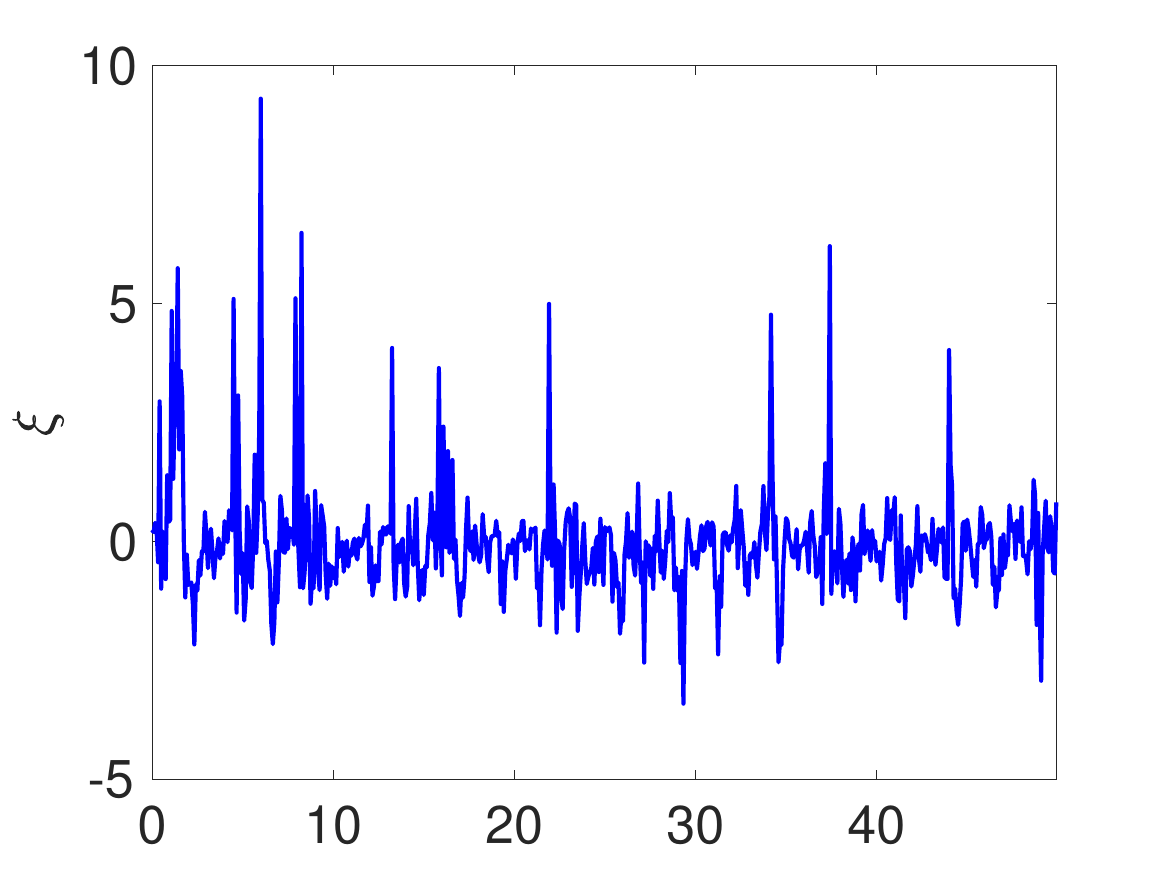}
    \put(14,73){\small (b)}
    \end{overpic}
	\caption{\textbf{Time series of multiplicative Gaussian noise and of CON noise over 50 years} (a) Multiplicative Gaussian noise model with $\xi_T = \sigma_T(1+B{\rm{max}}(T_e,0)) \zeta_t + T_e^2 + cT_e^3$ with parameters from \cite{Vialard-Jin-McPhaden-et-al-2025:el} and Gaussian noise. (b) CON noise model with parameters as in Table~\ref{tab:para}. }
\label{fig:Vialard_CON}
\end{figure}

To illustrate to what degree the conditional noise model (CON) generates forcing signals that are more consistent with the observed WWB time series as shown in Figure~\ref{fig:obs}a, we show in Figure~\ref{fig:Vialard_CON} a random realization of the CON model and of the multiplicative Gaussian noise model with $\xi_T=\sigma_T(1+B{\rm{max}}(T_e,0)) \zeta_t + T_e^2 + cT_e^3$ with Gaussian noise $\zeta_t$ that is used in \cite{Vialard-Jin-McPhaden-et-al-2025:el}, representing a commonly used choice in RO models. The raw additive CON noise $\xi$ nicely produces sporadic bursts that are clearly separated from a noisy background. In contrast, the large amplitude signals supported by the multiplicative Gaussian noise model are less pronounced. 

Our aim is to reproduce the statistical behaviour of both WWBs and the resulting observed signatures of ENSO, including those associated with extreme El Ni\~no events. Each of our three noise models is calibrated to reproduce the empirical histogram of the observed NINO3 time series, its power spectrum, and its variance and skewness, all based on the 153-year-long period from 1870 until 2026 (i.e., 1,872 months). We summarize the model parameters in Table~\ref{tab:para}. For the conditional noise (CON), we examine the three month average of the SST to determine the persistence of a positive or negative SST anomaly, based on which the noise is switched between OU and CAM. For computational ease, in the RO model \eqref{eq:Te0}--\eqref{eq:hw0} the average over the past three months is calculated as the average over the last three monthly snapshots, rather than as an average over all time steps occurring during the past three months. To match the empirical histogram for the conditional noise model (\textbf{CON}), we find we need to use the nonlinear SST response \eqref{eq:gamma}, which in effect shifts the histograms to higher values of the SST, enhancing its asymmetry.

\begin{figure}[!htp]
\begin{overpic}[width=0.32\textwidth]{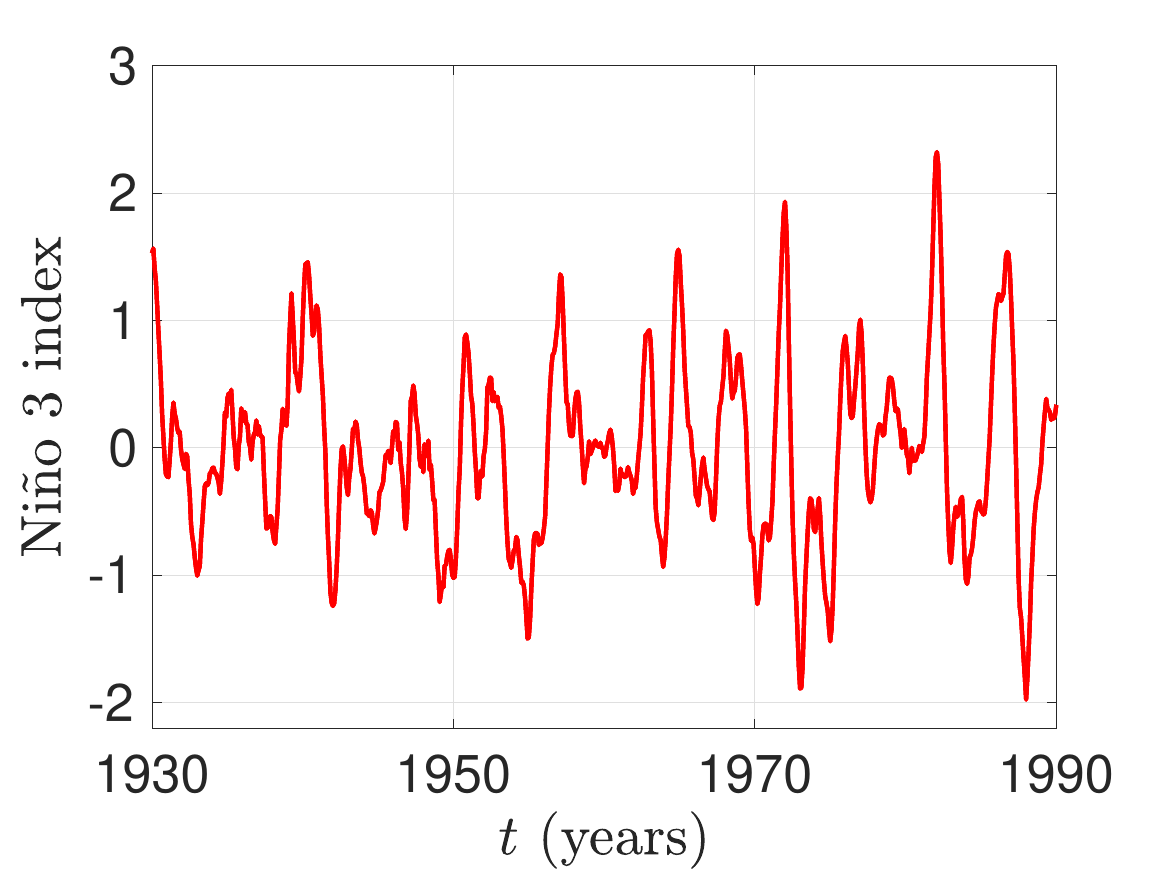}
    \put(14,73){\small (a)}
  \end{overpic}
  \begin{overpic}[width=0.32\textwidth]{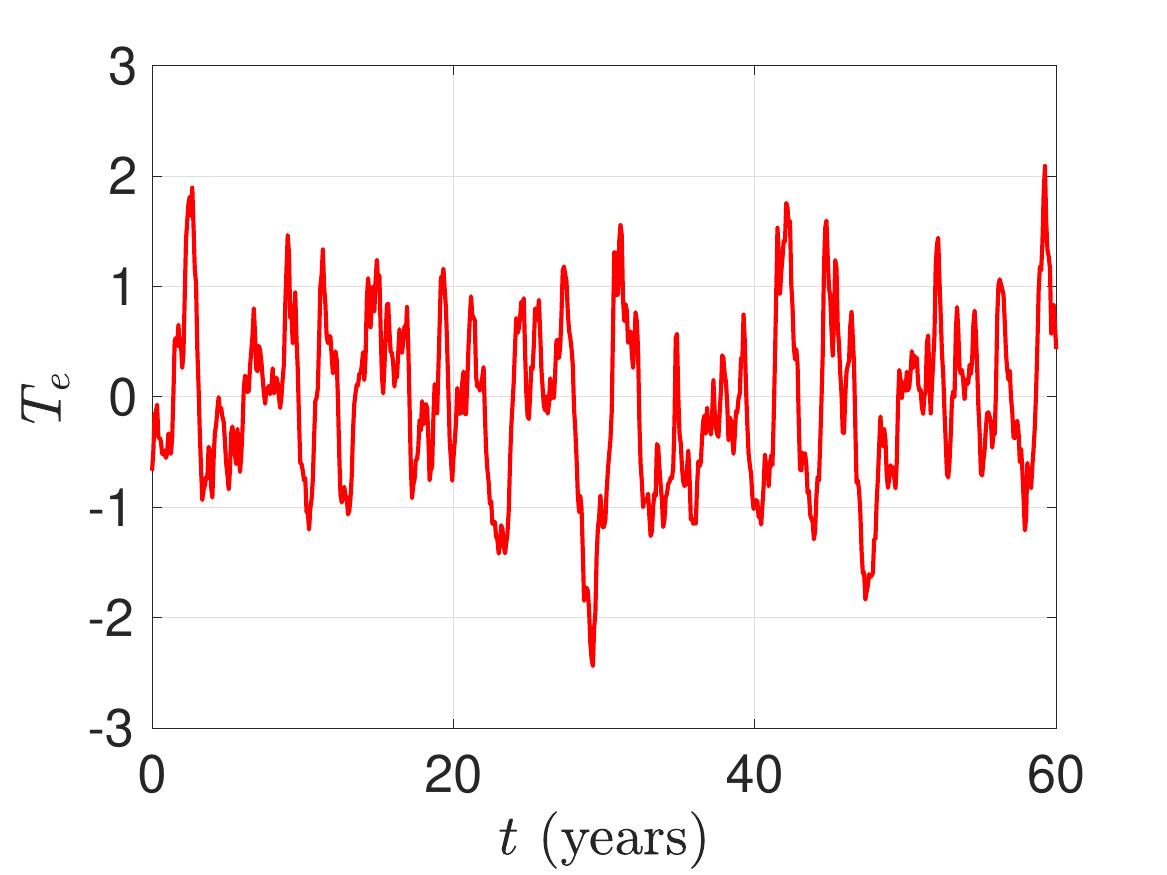}
    \put(14,73){\small (b)}
  \end{overpic}
  \\
\begin{overpic}[width=0.32\textwidth]{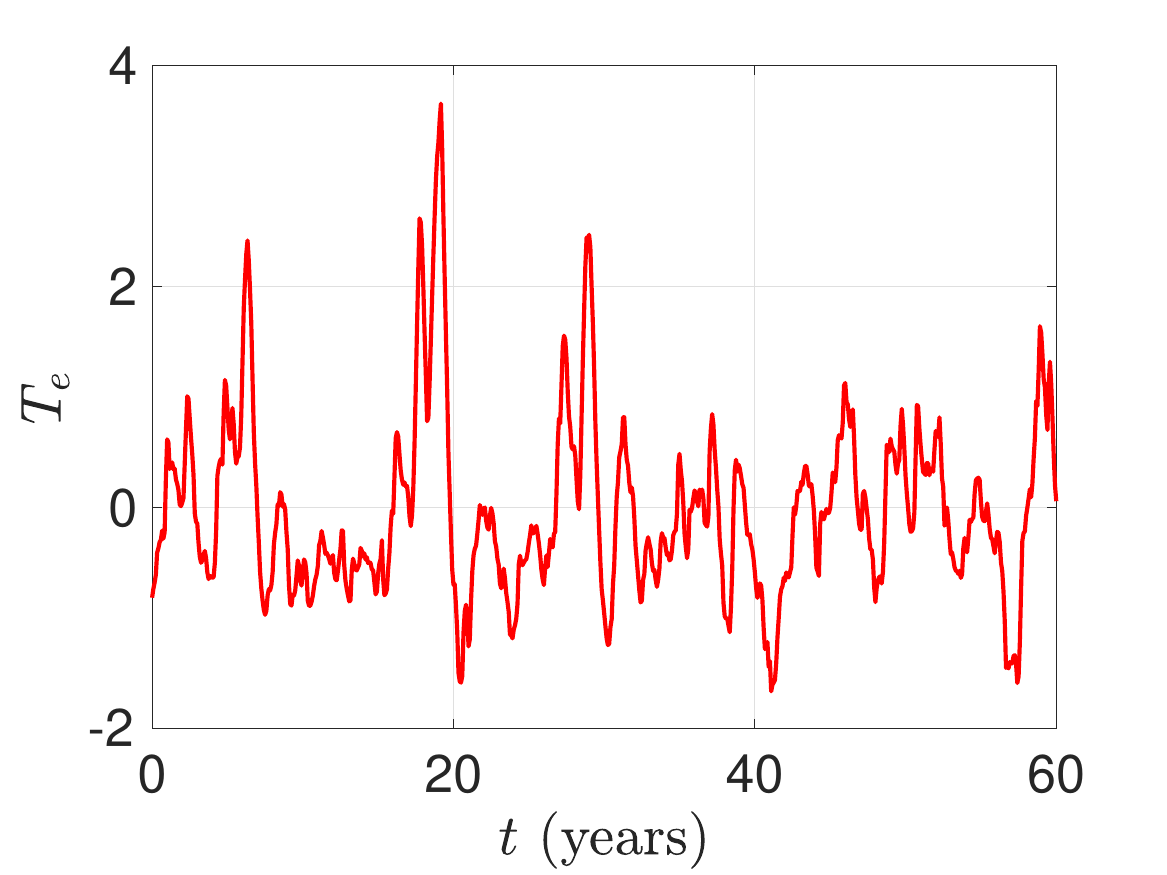}
    \put(14,73){\small (c)}
  \end{overpic}
\begin{overpic}[width=0.32\textwidth]{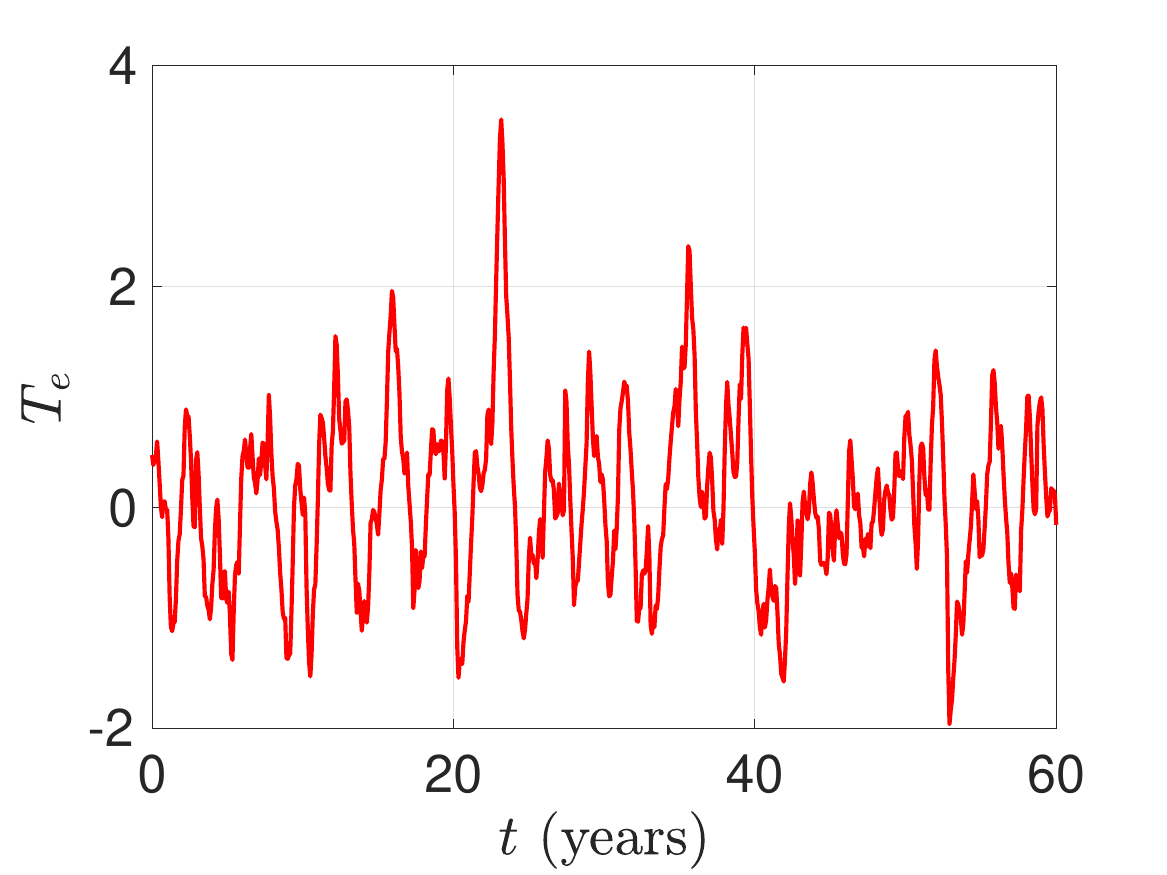}
    \put(14,73){\small (d)}
  \end{overpic}
\caption{\textbf{Comparison of typical SST time series for different noise models}. Time series of 60 years of the NINO3 index (a) are shown together with the SST $T_e$ obtained from the RO model \eqref{eq:Te0}--\eqref{eq:hw0} when driven by (b) OU, (c) CAM and (d) CON.}
\label{fig:TimeSeriesTe}
\end{figure}

Typical time series of the East Equatorial Pacific SST, $T_e$, are shown for the three noise models in Figure~\ref{fig:TimeSeriesTe}. We show some statistical characterizations of these time series below. Yet one might argue just based on these snapshots that multiplicative OU (Figure~\ref{fig:TimeSeriesTe}b) and the CON noise  (Figure~\ref{fig:TimeSeriesTe}d) seem to produce similar and reasonable ENSO simulations, as also supported by the statistical analyses below. The large sporadic peaks of the CAM noise as seen in Figure~\ref{fig:CAM}b give rise to large El Niño events and hence to a higher degree of asymmetry between El Niño and La Niña and a more skewed distribution compared to the OU process. We recall that we do not use the stochastic signal obtained from \eqref{eq:CAM} directly but cap the signal to be bounded and not to exceed a threshold of $\theta=3$.

\begin{figure}[!htp]
     \begin{overpic}[width=0.32\linewidth]{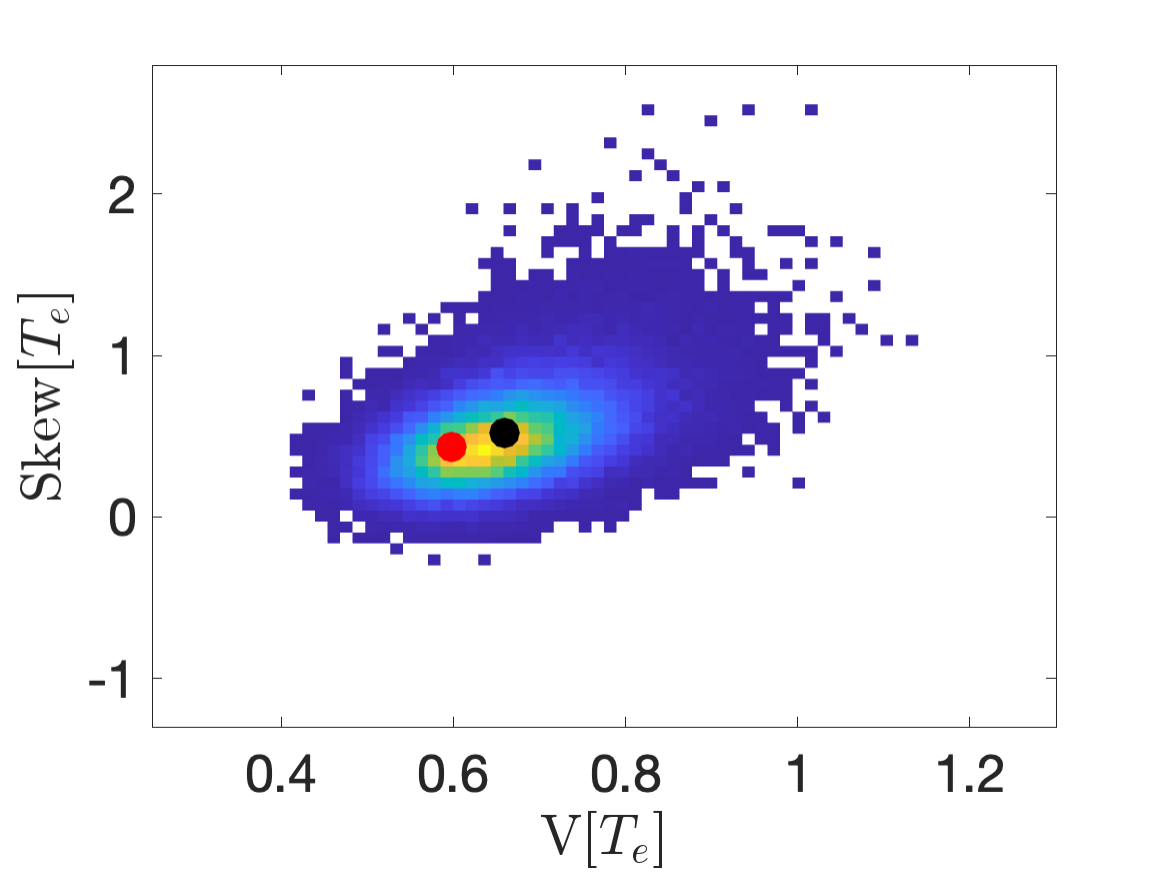}
    \put(14,73){\small (a)}
  \end{overpic}
     \begin{overpic}[width=0.32\linewidth]{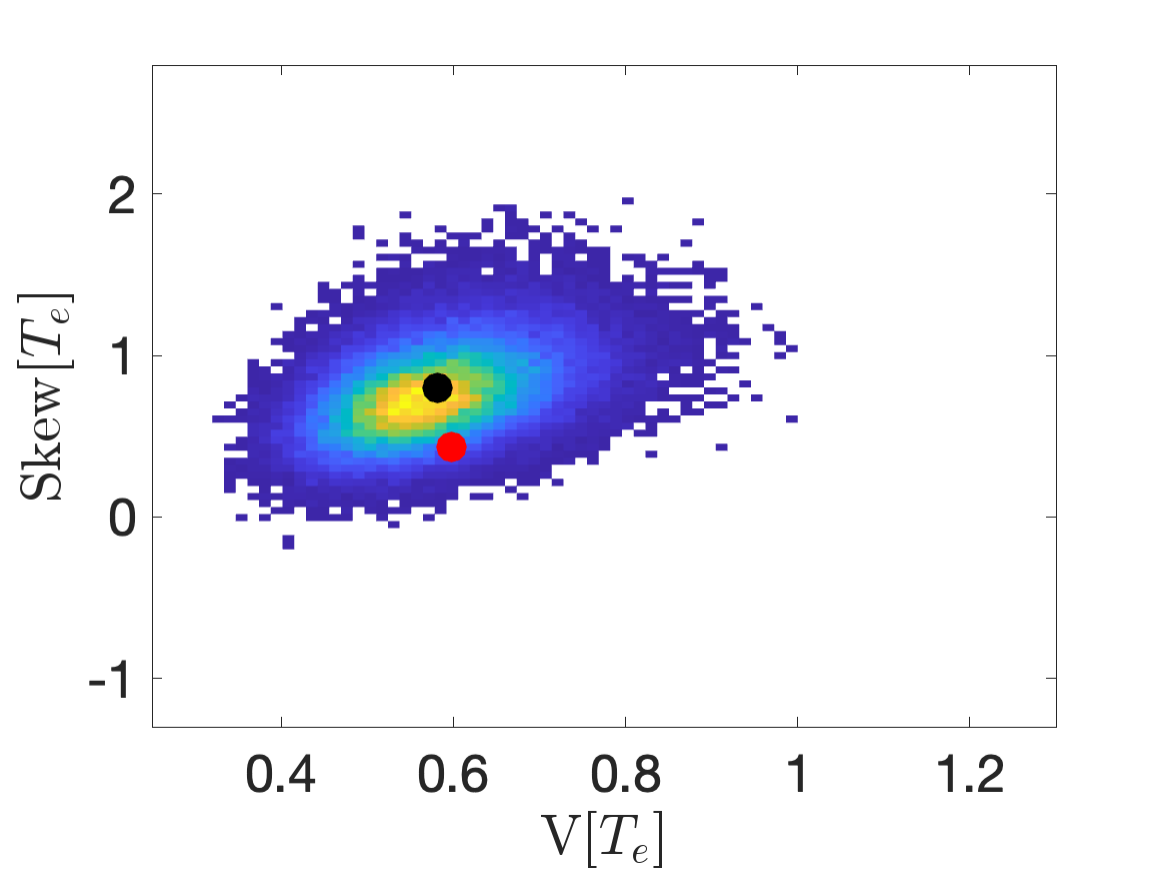}
    \put(14,73){\small (b)}
  \end{overpic}
     \begin{overpic}[width=0.32\linewidth]{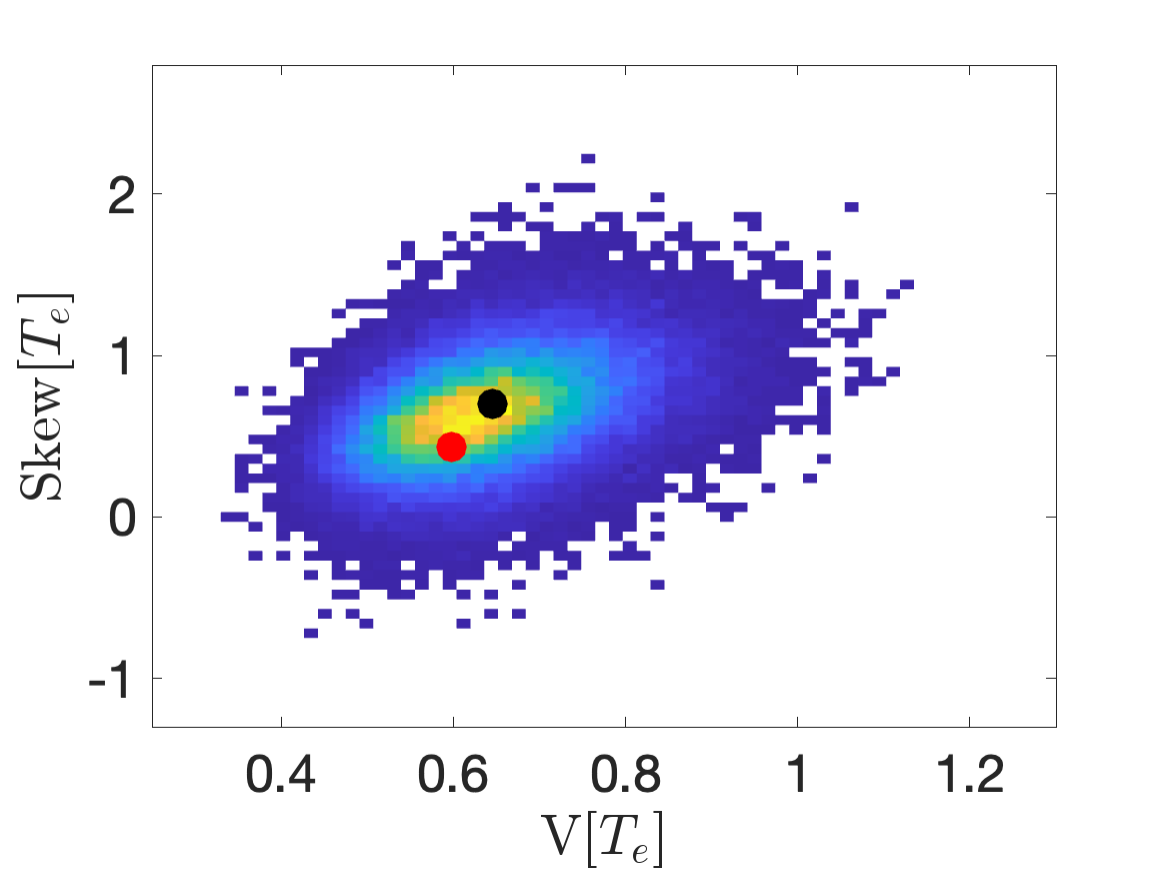}
    \put(14,73){\small (c)}
  \end{overpic}
\caption{\textbf{Histograms of the variance and skewness of the $T_e$ for different noise models.} The red dot demarcates the observed variance ${\rm{V}}[T_e]$ and skewness ${\rm{Skew}}[T_e]$ from the NINO3 index. The black dot is the average of a $10^8$ month-long simulation of the recharge oscillator model \eqref{eq:Te0}--\eqref{eq:hw0} when driven by  
(a) OU, (b) CAM, (c) CON.}
\label{fig:VarSkewHisto}
\end{figure}

Figure~\ref{fig:VarSkewHisto} shows a contour plot of the empirical 2D histogram of the variance and skewness for each 1,872-month-long segment of a $10^8$ months-long simulation of the SST $T_e$ for the RO model \eqref{eq:Te0}--\eqref{eq:hw0}, together with the mean of the variance and skewness on the observed NINO3 index. Note that the nonzero skewness for the additive CAM noise is entirely generated by the CAM noise process, which favours positive amplitudes with $\beta=0.8075$. Hence, CAM noise is capable of generating ENSO's asymmetry of having stronger El Ni\~no events than La Ni\~na events without any multiplicative noise or deterministic nonlinearities.

\begin{figure}[!t]
  \begin{overpic}[width=0.32\linewidth]{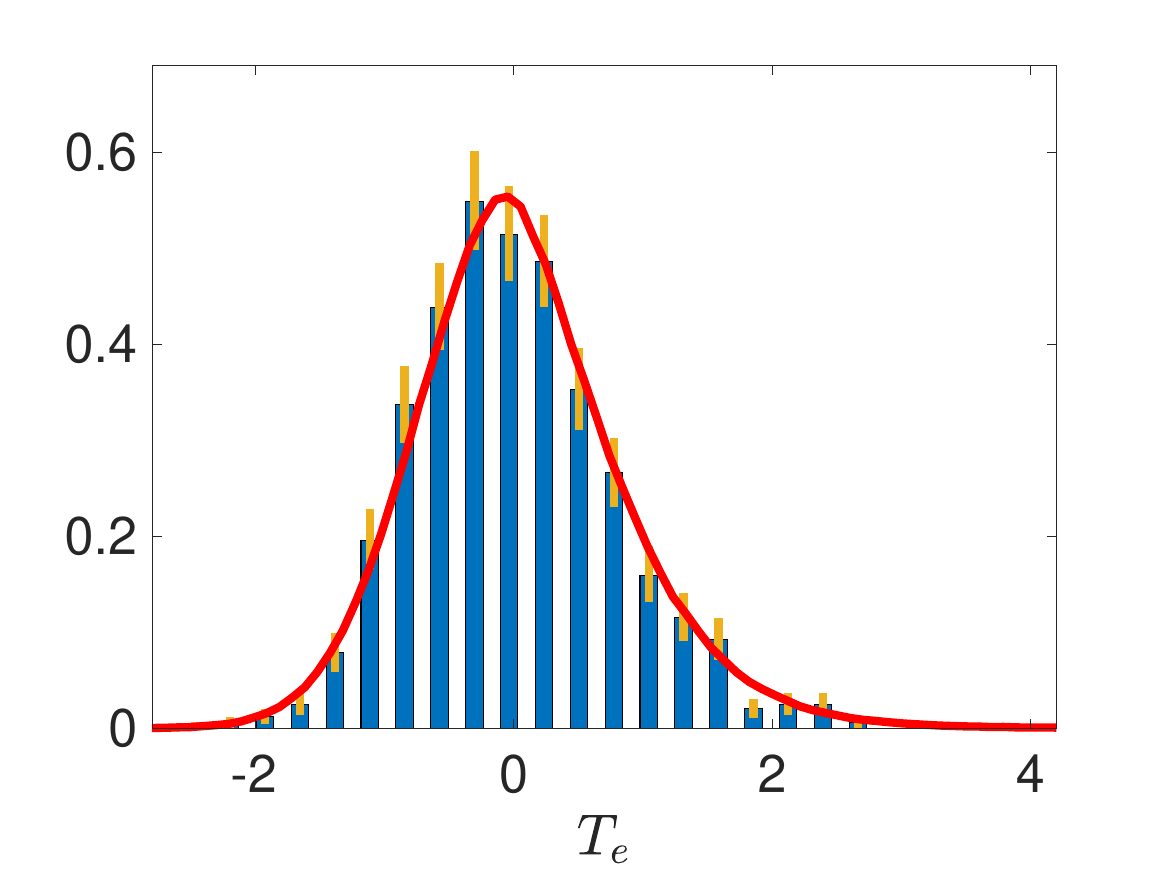}
    \put(14,73){\small (a)}
  \end{overpic}
  \begin{overpic}[width=0.32\linewidth]{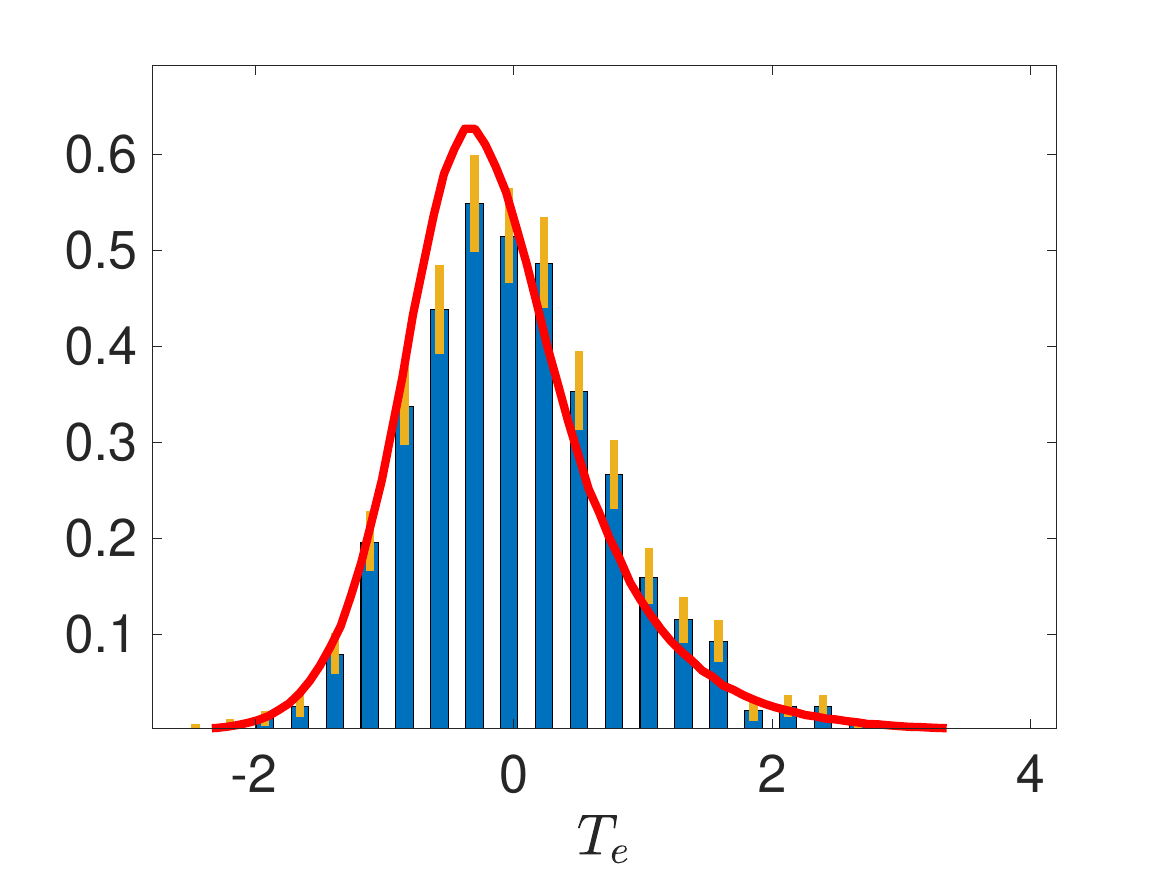}
    \put(14,73){\small (b)}
  \end{overpic}
  \begin{overpic}[width=0.32\linewidth]{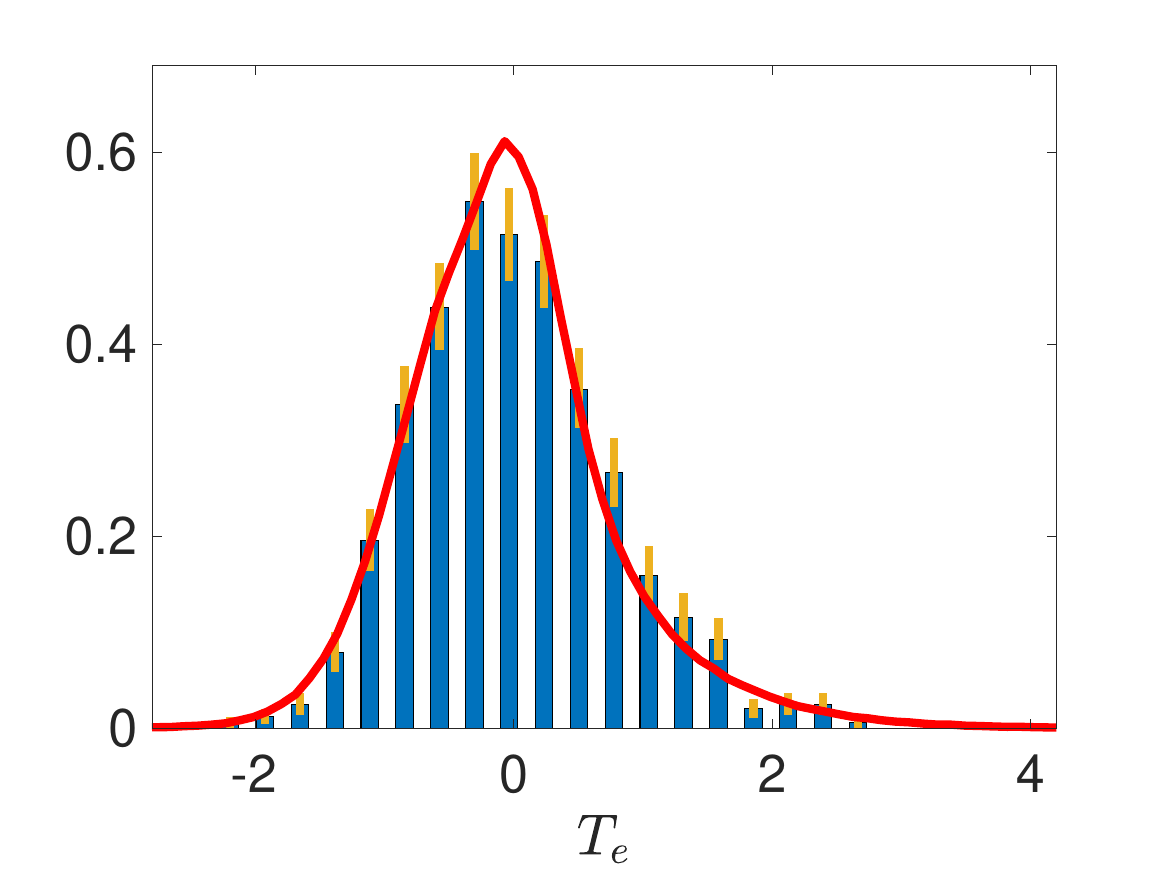}
    \put(14,73){\small (c)}
  \end{overpic}\\[0.5em]
  \begin{overpic}[width=0.32\linewidth]{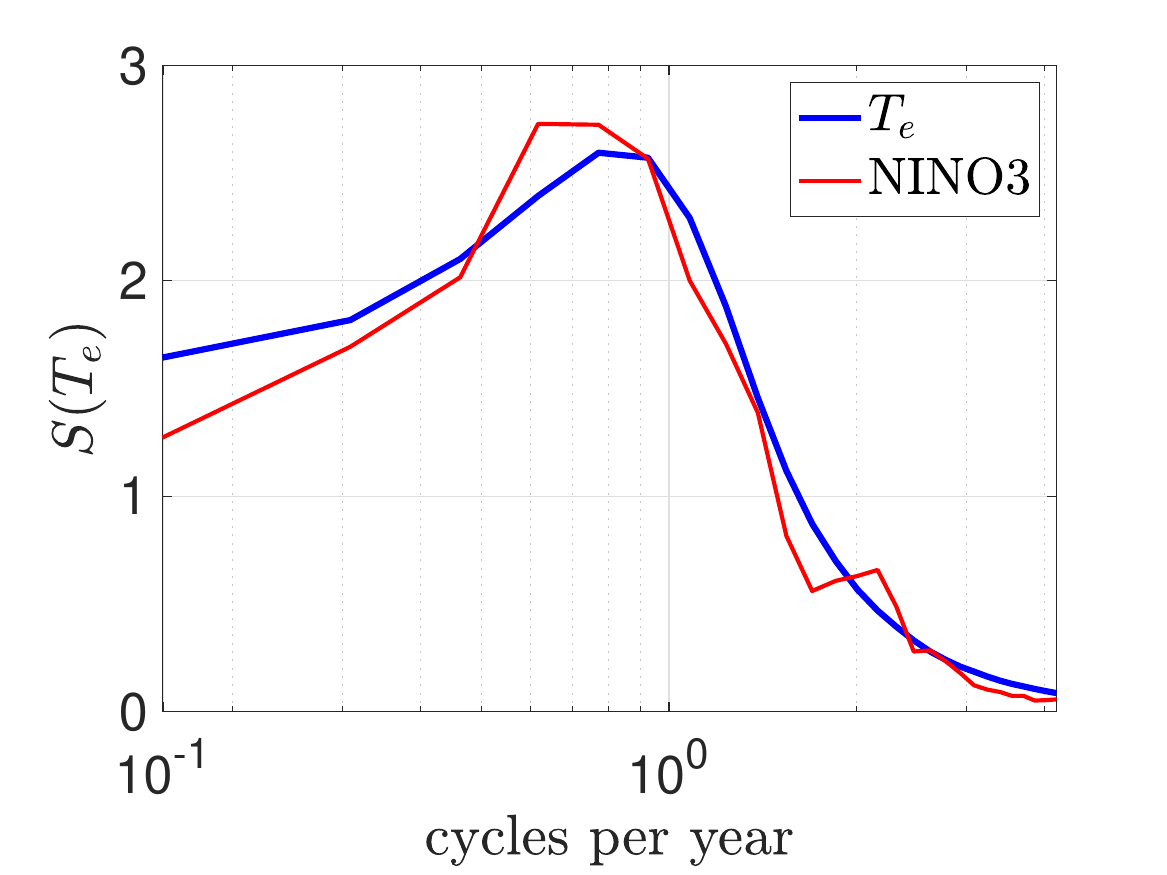}
    \put(14,73){\small (d)}
  \end{overpic}
  \begin{overpic}[width=0.32\linewidth]{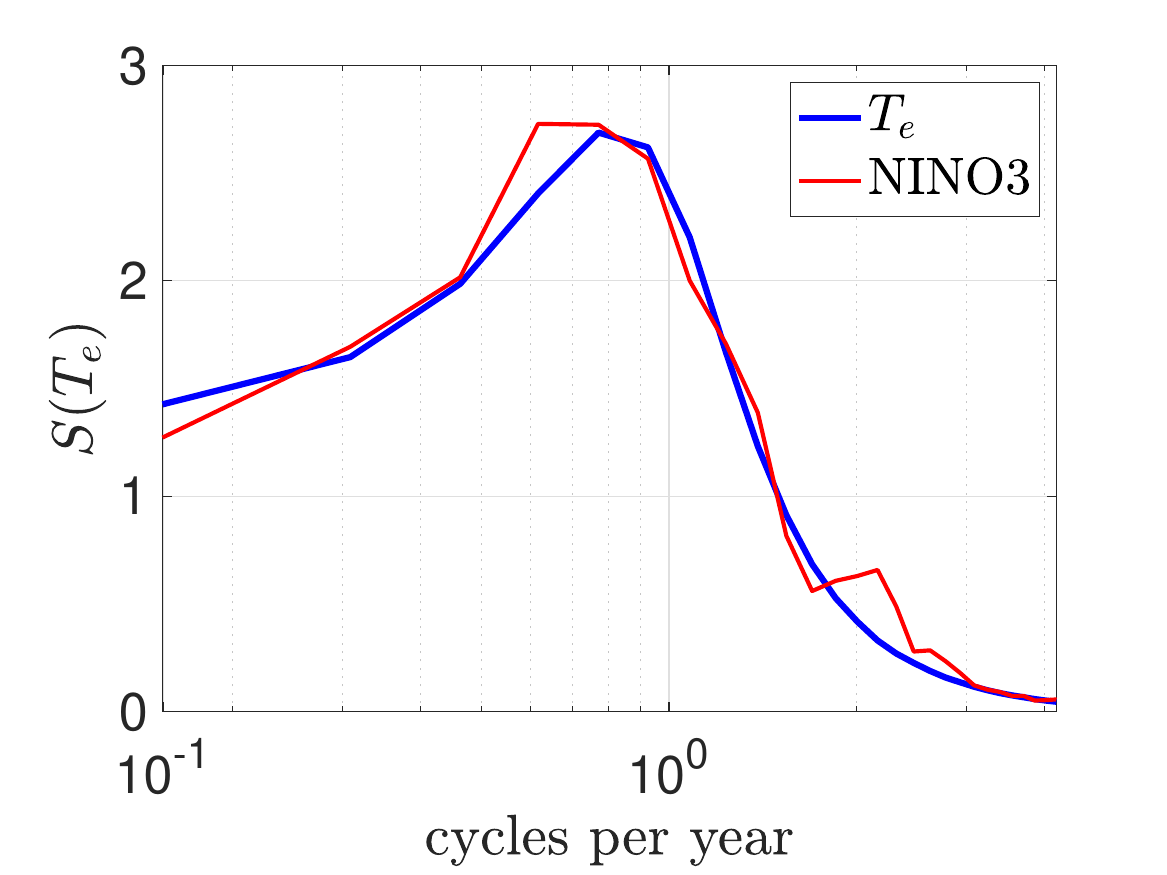}
    \put(14,73){\small (e)}
  \end{overpic}
  \begin{overpic}[width=0.32\linewidth]{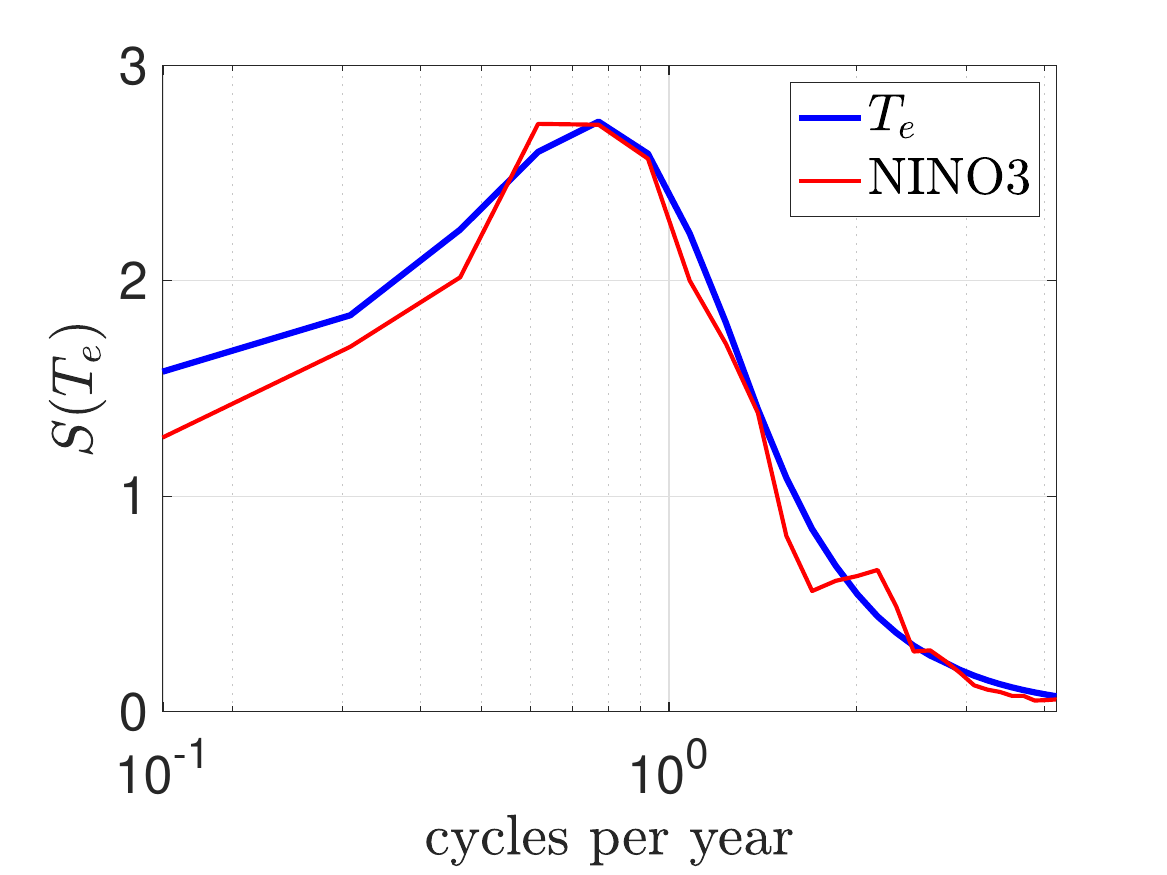}
    \put(14,73){\small (f)}
  \end{overpic}
  \caption{\textbf{Comparison of histograms and power spectra of observations and the RO model driven by the different noise models}. Top panels: Histograms of the observed NINO3 index from 1870 until 2026 with a monthly resolution and with the seasonal cycle removed (blue bars) and of the solution of the RO model (\ref{eq:Te0})--(\ref{eq:hw0}) shown by continuous curves, for the three noise models, (a) OU, (b) CAM, (c) CON. We show the 5th and 95th percentile confidence levels for the NINO3 index data as bars (beige) estimated from 10,000 bootstrapped samples. Bottom panels: Power spectra $S(T_e)$ of the observed NINO3 index from 1870 until 2026 with a monthly resolution and with the seasonal cycle removed (blue) and of the solution of the RO model (\ref{eq:Te0})--(\ref{eq:hw0}) for the three noise models, (d) OU, (e) CAM, (f) CON. A Welch window of $16$ years was employed.}
\label{fig:HistoTeComp}
\end{figure}

Figure~\ref{fig:HistoTeComp} shows the corresponding histograms for the SST $T_e$ and the power spectrum together with the corresponding curves for the observed NINO3 index. It is seen that all three noise models are capable of reproducing the global observed ENSO statistics reasonably well. To quantify to what degree each noise model can reproduce these statistics, we have compared several diagnostics. We performed a two-sample Kolmogorov-Smirnov test to check if any of the RO models generates SST data that could have been drawn from the distribution implied by the observational NINO3 index data. The null hypothesis was rejected for all three noise models, with CAM performing worst, as to be expected, as CAM noise is designed to generate only extreme warming events, rather than a wide range of events. The $p$-value of the OU model and the CON model are comparable with $p=0.0114$ and $p=0.0103$, respectively. We further estimated the Wasserstein distance between the synthetic $T_e$ model output and the NINO3  index, which measures the difference between their respective distributions. The OU model and the CON model again exhibit similar Wasserstein distances from observations, of 0.0552 and 0.0569, respectively. The data generated by the CAM noise exhibits again a larger Wasserstein distance from observations, of 0.1234, which is to be expected as these data are heavy tailed with several extreme events. To quantify the differences in the generated power spectra, we used a simple mean-square error. Here all three models show roughly the same error with 0.1344, 0.1091 and 0.115 for OU, CAM and CON, respectively. We remark that we did not optimize the equation and noise parameters to produce optimal fits with the observations but rather used a visual assessment. 

In summary, all three noise models are capable of producing the global bulk statistics of the observations, measured by the histograms, power spectra, and the variance and skewness. The CAM noise model is designed to generate extreme events and hence performs worse on the histograms. However, our aim for the noise model here is to not only fit the global statistical features of ENSO, but also to reproduce some of the dynamical features and signatures associated with WWBs.


\begin{figure}[!htp]
    \centering
    \begin{overpic}[width=0.32\linewidth]{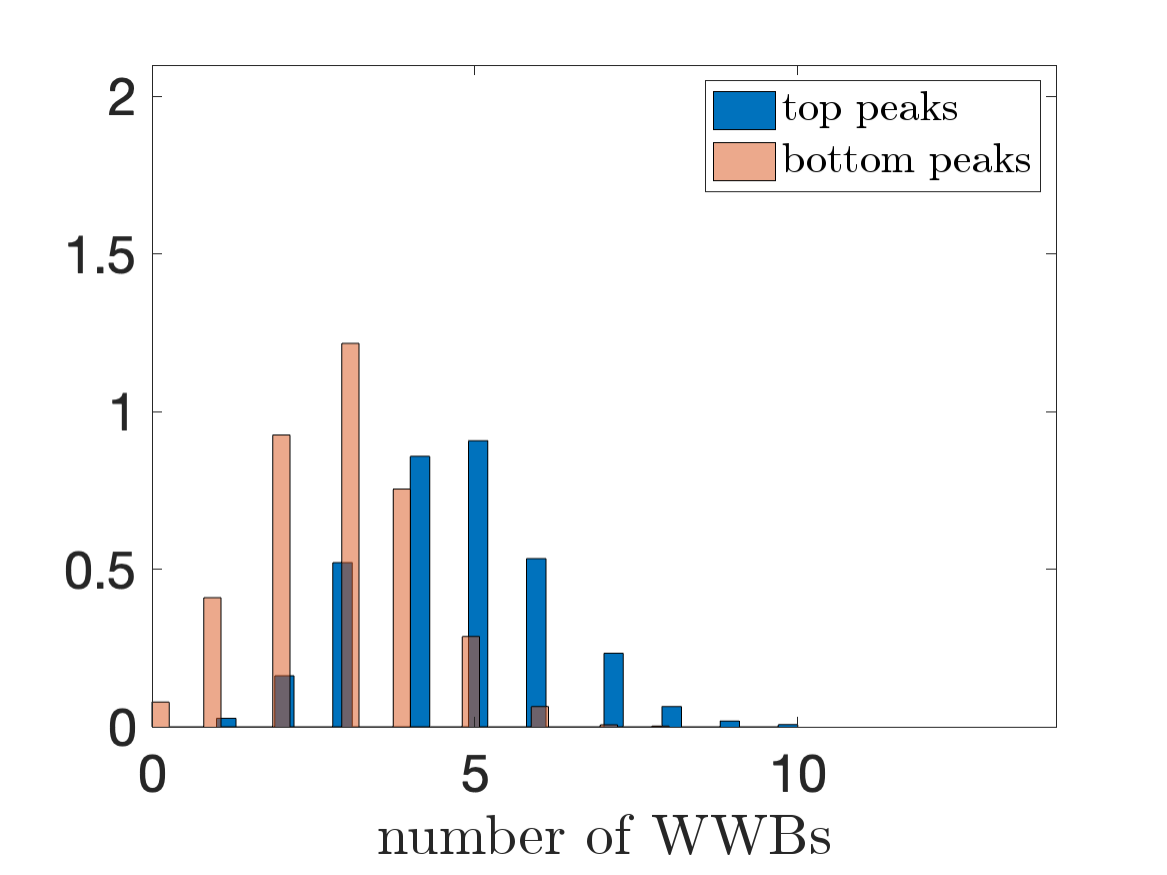}   
    \put(14,73){\small (a)}
  \end{overpic}
    \begin{overpic}[width=0.32\linewidth]{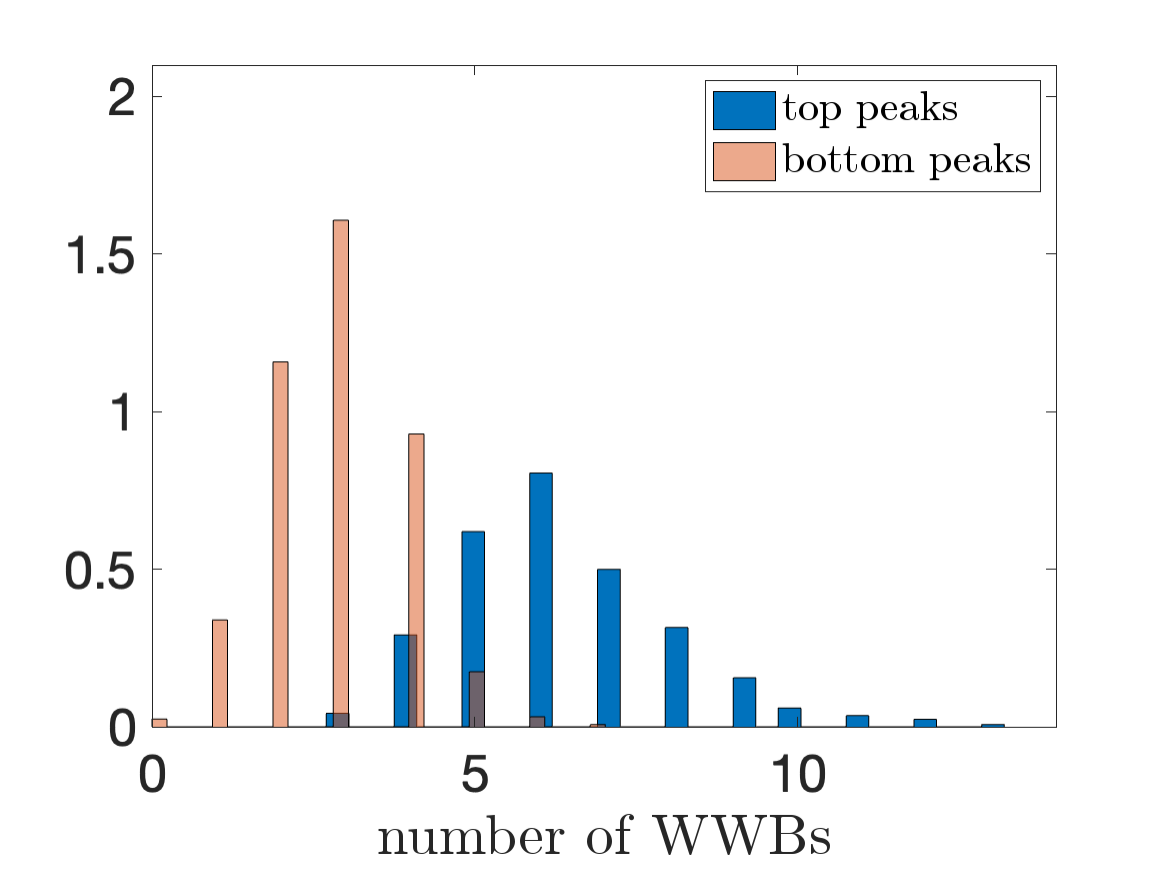}   
    \put(14,73){\small (b)}
  \end{overpic}
    \begin{overpic}[width=0.32\linewidth]{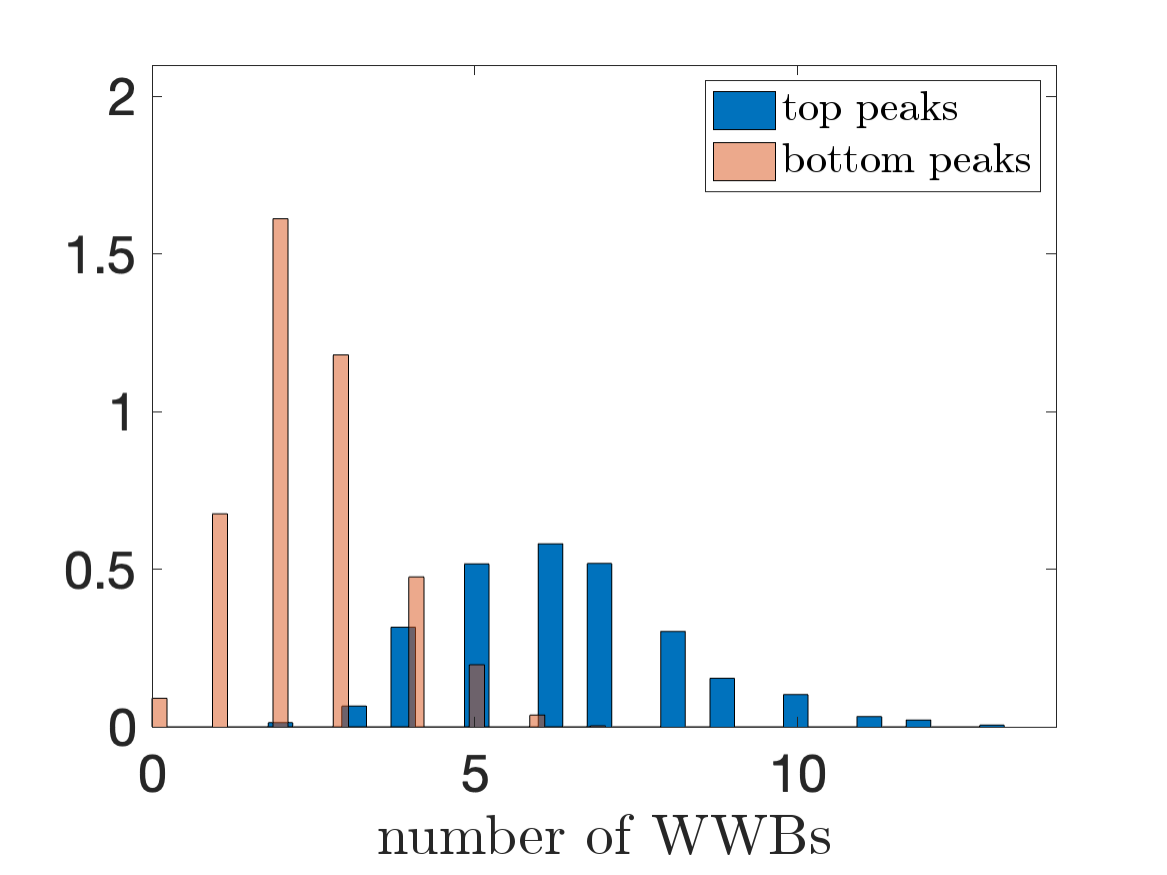} 
    \put(14,73){\small (c)}
  \end{overpic}
\caption{\textbf{Occurrence of WWBs before peak events for the different noise models.} Empirical histogram of the number of large noise amplitude events with $\xi>10/\nu_0$ occurring in the 12 months preceding a warm event in the RO model (\ref{eq:Te0})--(\ref{eq:hw0}). Shown are the empirical histograms for the largest and the smallest fifths of El Niño events (with a duration of at least 4 months). Parameters as in Figure~\ref{fig:VarSkewHisto}. (a) OU, (b) CAM, (c) CON.}
\label{fig:numb_WWB}
\end{figure}

To better distinguish the capability of the respective noise models to reproduce the effect of large WWBs on the dynamics, we now seek more fine-grained dynamical signatures associated with large-amplitude El Ni\~no events.  Observed strong El Niño events are accompanied by a sequence of WWBs \citep{Chiodi-Harrison-Vecchi-2014:subseasonal,Liang-Fedorov-2021:linking}. This is part of the positive feedback of warmer SSTs promoting the probability of the occurrence of WWBs, and WWBs intensifying the SST warming \cite[e.g.,][]{Tziperman-Yu-2007:quantifying, Eisenman-Yu-Tziperman-2005:westerly, Gebbie-Eisenman-Wittenberg-et-al-2007:modulation}. Figure~\ref{fig:numb_WWB} shows the number of months in which WWBs occurred during the 12-month period preceding an El Niño event. Strong WWBs are defined as large noise events with $\xi>10/\nu_0$; for CON we choose the smaller of the two values for $\nu_0$. We show results for the 20\% largest and for the 20\% smallest El Niño events (out of a long simulation of a total of $10^6$ months). Consistent with the observation of strong El Niño events co-occurring with several WWBs, the temperature $T_e$ of the strongest El Niño events is accompanied by a much larger number of WWBs for CAM and CON noise compared to small El Niño events (Figures~\ref{fig:numb_WWB}b,c). This effect is still observable to some degree for the multiplicative OU process (Figure~\ref{fig:numb_WWB}a), albeit to a much smaller degree.\\

\begin{figure}[!htp]
    \centering
\includegraphics[width=0.5\linewidth]{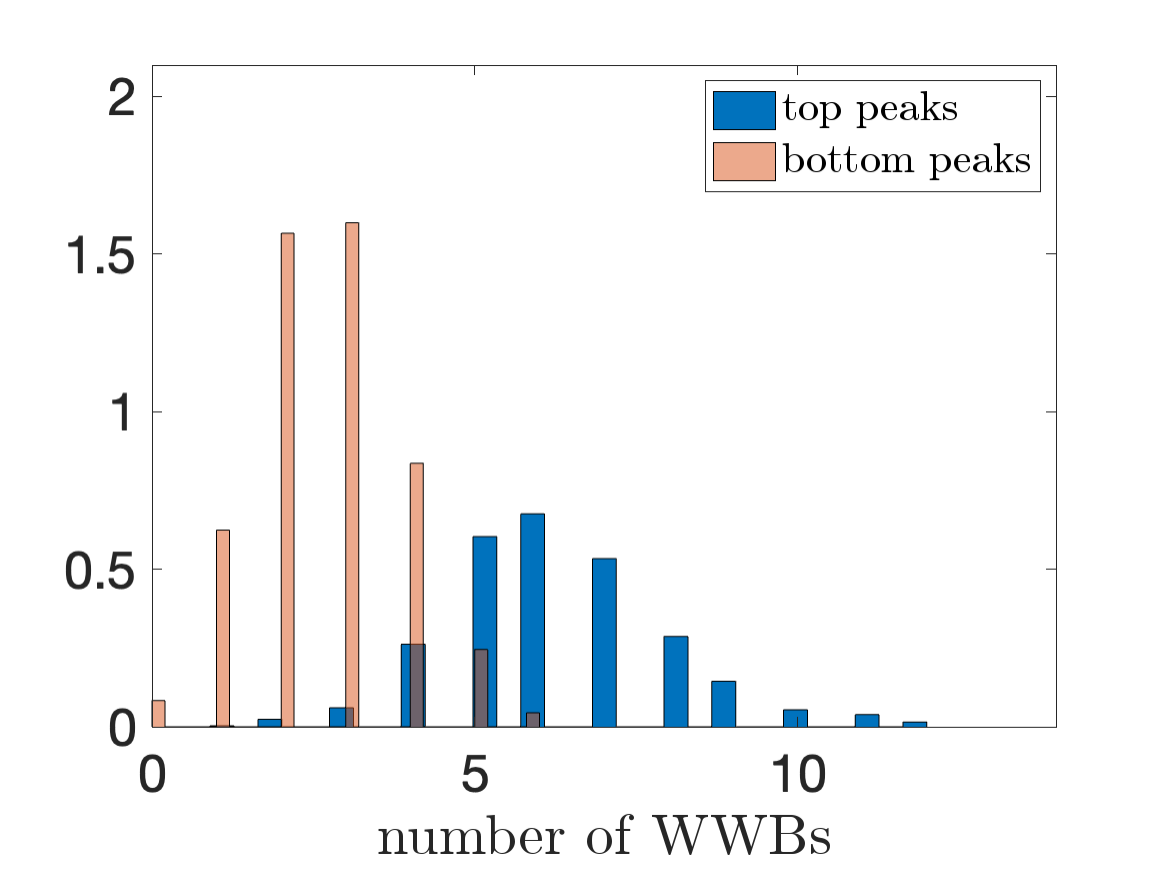} 
\caption{\textbf{Showing that the asymmetric thermocline response is not responsible for the increased number of WWBs prior to extreme El Ni\~no events for CON (Fig.~\ref{fig:numb_WWB}c).} Empirical histogram of the number of large noise amplitude events with $\xi>10/\nu_0$ occurring in the 12 months preceding a warm event for the RO model (\ref{eq:Te0})--(\ref{eq:hw0}) driven by a conditional noise model. All parameters are as for CON but with a symmetric response $\gamma_1=\gamma_2=0.08$ and adjusted noise amplitudes $\nu_0=17.5$ for the OU noise component and $\nu_0=5.4$ for the CAM component.}
\label{fig:CON_sym}
\end{figure}

Recall that, contrary to the OU and the CAM noise model, the CON noise model involves an asymmetric response of the SST with $\gamma_1\neq \gamma_2$. An obvious question is, if the observed behaviour depicted in Figure~\ref{fig:numb_WWB}c is due to this asymmetric response rather than to the proposed conditional CAM noise model. To clarify this, we show in Figure~\ref{fig:CON_sym} the empirical histogram of the number of large noise amplitude events occurring during warm events for a conditional noise with a symmetric response (and adjusted noise amplitudes $\nu_0$ to match the observed power spectrum as well as the implied variance and skewness of $T_e$). It is clearly seen that the conditional noise model with a symmetric thermocline/SST response also exhibits the characteristic behaviour observed for real WWBs. The asymmetric response is still required, however, to allow for a better approximation of the empirical histogram of the temperature (cf. Figure~\ref{fig:HistoTeComp}c); a symmetric response leads to a histogram shifted to smaller temperatures (not shown).

%

%
%%%%%%%%%%%%%%%%%%%%%%%%%%%%%%%%%%%%%%%%%%%%%%%%
%

\section{Conclusions}
\label{sec:disc}

We explored three different noise models for representing the effects of WWBs on ENSO in an RO model. One is multiplicative \textbf{OU} noise, a choice often used in the literature for this purpose. Another is based on correlated additive and multiplicative (\textbf{CAM}) noise \citep{Sura-Sardeshmukh-2008:global, Sardeshmukh-Sura-2009:reconciling, Penland-Sardeshmukh-2012:alternative, Sardeshmukh-Penland-2015:understanding} noise, characterized by the occurrence of sporadic large peaks, resembling realistic WWB events (cf. Figure~\ref{fig:obs}a). And a final one that conditionally switches from OU for negative NINO3 to CAM for positive NINO3 (\textbf{CON}), where this conditional switching is motivated by the observation that WWBs respond to the SST \citep{Yu-Weller-Liu-2003:case, Tziperman-Yu-2007:quantifying}.

All three models were able to explain the global ENSO statistics such as the empirical histogram, the power spectrum, and the variance and the skewness of the observed NINO3 time series. The various noise models, however, give rise to different noise characteristics that are not all appropriate for representing WWBs. The point we are making here is that, from an understanding point of view, it is not sufficient for the noise model to produce ENSO's characteristics, but the noise itself should resemble the observed WWB characteristics. Furthermore, characteristics such as having more WWBs preceding large El Ni\~no events should be accounted for as well. Our results suggest OU noise may be an appropriate forcing leading to normal El Ni\~no events of small and moderate amplitudes. Whereas we find that CAM noise is better suited to generate high-amplitude events, accounting for a larger number of WWBs preceding them. Our proposed conditional noise combines OU and CAM noise models to construct a noise model that accounts for both small and large amplitude El Ni\~no events, allowing for a better dynamical representation of WWBs. In particular, CON noise generates a noise forcing that more faithfully represents the sporadic nature of WWBs when compared to additive and multiplicative Gaussian noise models. Moreover, CON noise reproduces the observed dynamical signature that extreme El Niño events are preceded by several WWB events in the 12 months preceding the peak.

Further exploration of this idea using more realistic climate models seems an appropriate future direction.

%
%%%%%%%%%%%%%%%%%%%%%%%%%%%%%%%%%%%%%%%%%%%%%%%%
%

%\codeavailability{TEXT} %% use this section when having only software code available
%\dataavailability{TEXT} %% use this section when having only data sets available

%\codedataavailability{The code is available from the authors upon reasonable request. Data were obtained from publicly available repositories} %% use this section when having data sets and software code available

%
%%%%%%%%%%%%%%%%%%%%%%%%%%%%%%%%%%%%%%%%%%%%%%%%
%

\section*{Appendix: Model parameters}

\begin{table}[!htp]
    \centering
    \setlength{\tabcolsep}{2.5pt} % default is 6pt, decrease to reduce width
    \renewcommand{\arraystretch}{1.2} % slightly increase row spacing to improve readability
    % \scriptsize % or \footnotesize for less reduction
    \begin{tabular}{|c|c|c|c|}
         %\multicolumn{4}{l}{} \\[0.5ex] 
         \hline
         \textbf{} & \textbf{multiplicative OU} & \textbf{additive CAM} & \textbf{conditional OU/CAM}\\  
         \hline
         %\noalign{\vskip 1.0ex}
         %\hline       
         $r$  & $0.28$ & $0.25$ & $0.17$ \\ \hline
         $\varepsilon$ & $1/2.7$ & $1/2.7$ & $1/2.7$ \\ \hline
         $a$ & $0.41$ & $0.41$ & $0.33$ \\ \hline
         $b$ & $14$ & $14$ & $14$ \\ \hline
         $\gamma_+$ & $0.088$ & $0.09$ & $0.08$ \\ \hline
         $\gamma_-$ & $0.088$ & $0.09$ & $0.0728$ \\ \hline   
         $\nu_0$ & $20$ & $6$ & $21$/$6.5$ \\ \hline
         $\nu_1$ & $0.2791$ & $0$ & $0$ \\ 
         %\hline
         %\noalign{\vskip 1.0ex}
         \hline
         $c_1$ & $-1.4$ & $-1.22$  & $-1.4$/$-1.19$ \\ \hline
         $c_2$ & $0$ & $1.14$ & $0$/$1.12$ \\ \hline
         $c_3$ & $0$ & $0.65$ & $0$/$1.2$ \\ \hline
         $c_4$ & $0.7$ & $0.8$ & $0.7$/$1.1$ \\ \hline
     \end{tabular}
     \vskip 0.3cm
    \caption{Parameters for the RO model \eqref{eq:Te0}--\eqref{eq:hw0} and the noise models \eqref{eq:CAM}. Here the units are as follows: $[r]=1/{\rm{month}}$, $[\varepsilon]=1/{\rm{month}}$, $[a]=m^3/N$, $[b]=1/K$, $[\gamma_{\pm}]=K/m$, $[\nu_0]=1$, $[\nu_1]=1/K$.}
    \label{tab:para}
\end{table}

%
%%%%%%%%%%%%%%%%%%%%%%%%%%%%%%%%%%%%%%%%%%%%%%%%
%

\section*{Acknowledgments}
AF and ET were funded by the Department of Energy (DOE) Office of Science Biological and Environmental Research grant DE-SC0023134. ET is also funded by the Harvard Dean’s Competitive Fund for Promising Scholarship, and thanks the Weizmann Institute of Science for its hospitality during parts of this work.

%% REFERENCES

%%%\bibliographystyle{copernicus}
\bibliography{../export}

\end{document}